\documentclass[aps,prd,preprint,nofootinbib,groupedaddress]{revtex4}
\usepackage{amsmath,amssymb,amsbsy,color} 
\usepackage{graphicx}
\usepackage{psfrag}

\newcommand{\nn}{\nonumber}

 \newcommand{\pslash}{p\kern-1ex /}
\newcommand{\lslash}{l\kern-1ex /} \newcommand{\sslash}{s\kern-1ex /}
\newcommand{\Dslash}{{\cal D}\kern-1.5ex /}

\newcommand{\beqa}{\begin{eqnarray}}
\newcommand{\eeqa}{\end{eqnarray}} \newcommand{\be}{\begin{equation}}
\newcommand{\ee}{\end{equation}} \newcommand{\bea}{\begin{eqnarray}}
\newcommand{\eea}{\end{eqnarray}} \newcommand{\ba}{\begin{array}}
\newcommand{\ea}{\end{array}}

\newcommand{\pref}[1]{(\ref{#1})}

\newcommand{\Lag}[1]{{\cal L}_{\rm {#1}}} \newcommand{\Lagno}{{\cal
    L}} \newcommand{\psibar}{\overline{\psi}}
\newcommand{\Ldag}{L^{\dagger}} \newcommand{\Rdag}{R^{\dagger}}
\newcommand{\psiL}{\psi_{L}} \newcommand{\psiR}{\psi_{R}}
\newcommand{\psibarL}{\overline{\psi}_{L}}
\newcommand{\psibarR}{\overline{\psi}_{R}}  
\newcommand{\Tr}[1]{\langle {#1} \rangle}
\newcommand{\csw}{\overline{c}_{{\rm SW}}}
\newcommand{\ca}{\overline{c}_{{\rm A}}}
\newcommand{\cv}{\overline{c}_{{\rm V}}}

\newcommand{\VC}[2]{V^{#1}_{\mu,{\rm #2}}}
\newcommand{\pimu}{\pi_{\mu}} 

\begin{document}

\preprint{UTHEP-581/HU-EP-09/02/SFB-CPP-09-25} \title{ Vector and
  Axial Currents in Wilson Chiral Perturbation Theory}

\author{$^{1,2}$Sinya Aoki, $^{3}$Oliver B\"ar and $^{4}$Stephen
  R. Sharpe}

\affiliation{ $^1$Graduate School of Pure and Applied Sciences,
  University of Tsukuba, Tsukuba 305-8571, Ibaraki Japan \\ $^2$Riken
  BNL Research Center, Brookhaven National Laboratory, Upton, NY
  11973, USA \\ $^3$Institute of Physics, Humboldt University Berlin,
  Newtonstrasse 15, 12489 Berlin, Germany\\ $^4$Physics Department,
  University of Washington, Seattle, WA 98195-1560, USA
  }

\date{\today}

%
\begin{abstract}
%
We reconsider the construction of the vector and axial-vector currents
in Wilson Chiral Perturbation Theory (WChPT), the low-energy effective
theory for lattice QCD with Wilson fermions.  We discuss in detail the
finite renormalization of the currents that has to be taken into
account in order to properly match the currents. We explicitly show
that imposing the chiral Ward identities on the currents does, in general,
affect the axial-vector current at O($a$).  As an application of our results
we compute the pion decay constant to one loop in the two flavor
theory. Our result differs from previously published ones.

\end{abstract}
\maketitle

\section{Introduction}
\label{sect:intro}

Chiral perturbation theory (ChPT) is widely regarded to be an
important tool for Lattice QCD. It provides analytic guidance for the
chiral extrapolation of the lattice data obtained at quark masses
heavier than in nature.  Standard ChPT, as formulated in
Refs.\ \cite{Gasser:1983yg,Gasser:1984gg}, is based on the symmetries
and the particular symmetry breaking of continuum QCD. 
The generalization to Lattice QCD with Wilson fermions,
taking into account the explicit chiral symmetry breaking of Wilson's fermion
discretization~ \cite{Wilson:1974sk},
was given in Ref.~\cite{Sharpe:1998xm}.
The resulting low-energy effective theory, often
called Wilson chiral perturbation theory (WChPT), is a double
expansion in the quark mass and the lattice spacing, the two
parameters of explicit chiral symmetry breaking.

Massless continuum QCD is invariant under various non-singlet chiral
transformations. This invariance implies the existence of conserved
currents (which are obtained by the Noether theorem) and various
chiral Ward identities. ChPT is constructed in such a way that these
Ward identities are correctly reproduced, order by order in the chiral
expansion. And since conserved currents do not renormalize it is
straightforward to maintain the normalization of the currents.

The construction of WChPT is slightly more complicated compared to
continuum ChPT.  Due to the explicit breaking of chiral symmetry by
the Wilson term there does not exist a conserved axial-vector current
for vanishing bare quark mass. And even though a conserved vector
current exists for degenerate quark masses, it is often not used in
practice. The local, non-conserved vector current is employed instead,
even though it requires the computation of a renormalization constant
$Z_{V}$. The renormalization constant $Z_{A}$ is also needed for the
axial-vector current.

The explicit breaking of chiral symmetry and the need for
renormalizing the currents raises the question how to construct the
effective currents in Wilson ChPT. The ``Noether link'' does not hold
anymore. Also the renormalization of the
currents has to be taken into account for a proper matching of the
effective theory to the fundamental lattice theory.

Some results concerning the currents can be found in the literature
\cite{Rupak:2002sm,Sharpe:2004ny},
but they are in conflict.
Reference~\cite{Rupak:2002sm} calculates the pion decay constant
using the current obtained by the naive Noether procedure as the axial-vector current
\cite{ShoreshPC}.  
Reference~\cite{Sharpe:2004ny} introduces source terms for the currents
as in continuum ChPT, 
and constructs the generating functional for correlation functions of
the currents. The resulting axial-vector current contains an additional O($a$)
contribution that is not present in the Noether current. Consequently,
the resultant $f_{\pi}$ contains an extra term and differs from that
of Ref.~\cite{Rupak:2002sm}.

Besides this discrepancy the issue of renormalization has not been
properly taken into account in either work. No particular
renormalization condition for the axial-vector current has been
imposed as is necessary for a proper matching of the currents. It has
been argued in Ref.\ \cite{Sharpe:2004ny} that the results for the
currents derived there should hold for any choice of lattice operators
which are correctly normalized in the continuum limit. However, the
validity of this expectation has not been shown so far.

In this paper we reconsider the construction and mapping of the vector
and axial-vector currents in WChPT to O($a$).\footnote{%
  Preliminary results have already been presented in
  Ref.\ \cite{Aoki:2007es}. Details have changed but the overall
  conclusions are unaltered.} 
We proceed in two steps.  First, we
write down the most general expressions for the currents which are
compatible with locality and the symmetries of the underlying lattice
theory. With this procedure we reproduce the results of
Ref.~\cite{Sharpe:2004ny} (which are more fully
justified in Ref.~\cite{SharpeNara}).  
In the second step we impose the
chiral Ward identities as particular renormalization conditions for
the currents. This choice, suggested in
Refs.\ \cite{Karsten:1980wd,Bochicchio:1985xa,Maiani:1986yj}, is
widely used in practice.  We find that this renormalization condition
does have an impact at O($a$) on the axial-vector
current. Consequently, our current differs from the ones in
Refs.\ \cite{Rupak:2002sm,Sharpe:2004ny}. As an application of our
results we finally compute the pion decay constant to one loop,
including the O($a$) correction to the chiral logarithm, which also
differs from the results in \cite{Rupak:2002sm,Sharpe:2004ny}.

This paper is organized as follows. In section \ref{sect:Currents} we
first summarize some definitions of the lattice theory with two
flavors of Wilson quarks, in particular the various vector and axial
vector currents used in numerical simulations. This is followed by the
Symanzik expansion of the currents close to the continuum limit. The
currents in the Symanzik effective theory are then mapped to their
counterparts in the chiral effective theory. Section
\ref{sect:Renormalization} discusses the renormalization of the vector
and axial-vector currents in the lattice theory and how this is
carried over to the effective theory. The results for the decay
constant are given in section \ref{sect:decayconstant}, followed by
some concluding remarks in section \ref{sect:conclusion}. Appendix
\ref{sourcemethod} is devoted to an alternative but equivalent
derivation of the currents based on the generating functional, while
appendix \ref{app_WTI} consists of details concerning the calculation
of $Z_A$ in the effective theory.

\section{Currents in WChPT}
\label{sect:Currents}

\subsection{Definitions in the Lattice theory}

We consider Lattice QCD with Wilson fermions on a hypercubic lattice
with lattice spacing $a$.  For simplicity we study $N_{f}=2$ quarks
with equal quark mass. The fermion action is of the form
\bea\label{eq:LatticeAction} S_{f} & = & S_{\rm W} + c_{\rm SW} S_{\rm
  clover}\,, \eea where the first part denotes the standard Wilson
action \cite{Wilson:1974sk} with bare quark mass $m_{0}$. We also
allow for a clover-leaf term with coefficient $c_{\rm SW}$. The details
of the gauge action are not important in the following so we leave it
unspecified.

It is common to use the local expressions for the vector and axial
vector currents in numerical simulations, 
\bea 
V_{\mu,{\rm Loc}}^{a}(x) & = & \overline{\psi}(x)
\gamma_{\mu}T^{a}\psi(x) \label{eq:LocV},
\\ 
A_{\mu,{\rm Loc}}^{a}(x) &= & \overline{\psi}(x)
\gamma_{\mu}\gamma_{5}T^{a}\psi(x).\label{eq:LocA} 
\eea 
The $T^{a}$ are the hermitian SU($N_{f}$) generators, normalized according to
$\mbox{tr}(T^{a}T^{b}) = \delta^{ab}/2$. In the case of $N_{f}=2$, the
one we are considering, this normalization corresponds to $T^{a} =
\sigma^{a}/2$, where $\sigma^{a}$ are the usual Pauli matrices.

For degenerate quark masses the fermion action \pref{eq:LatticeAction}
is invariant under SU($N_{f})$ flavor transformations. The associated
conserved vector current differs from \pref{eq:LocV} and reads
\cite{Karsten:1980wd} 
\bea 
V_{\mu,{\rm Con}}^{a}(x) & = &
\frac{1}{2}\bigg\{ \overline{\psi}(x+a\hat{\mu}
)\gamma_{\mu}T^{a}U_{\mu}(x) \psi(x) +
\overline{\psi}(x)\gamma_{\mu}T^{a}U_{\mu}^{\dagger}(x)
\psi(x+a\hat{\mu} ) \nonumber 
\\ 
& & +\, \overline{\psi}(x+a\hat{\mu}
)U_{\mu}(x) \psi(x) + \overline{\psi}(x)U_{\mu}^{\dagger}(x)
\psi(x+a\hat{\mu} ) \bigg\}
\label{eq:ConV}\,.  
\eea 
No conserved axial
vector current exists due to the explicit chiral symmetry breaking by
the Wilson term in $S_{\rm W}$.

In on-shell O($a$)-improved lattice theories with degenerate quarks
one defines improved currents by adding terms 
involving lattice derivatives to the local currents
\cite{Luscher:1996sc,Sint:1997jx}, 
\bea 
V_{\mu,{\rm Imp}}^{a}(x) & = &
(1+b_{\rm V} am)\left[
V_{\mu,{\rm Loc}}^{a}(x) + c_{\rm V}\frac{1}{2} 
(\nabla_{\nu}^{+} +
\nabla_{\nu})T_{\mu\nu}^{a}(x)\right]\,,
\label{eq:ImpV}
\\ 
A_{\mu,{\rm Imp}}^{a}(x) & = & 
(1+b_{\rm A} am)\left[
A_{\mu,{\rm Loc}}^{a}(x) + c_{\rm A}\frac{1}{2}
(\nabla_{\mu}^{+} + \nabla_{\mu})P^{a}(x)\right]\,,
\label{eq:ImpA}
\eea 
where
\begin{equation}
T^{a}_{\mu\nu,{\rm Loc}}(x) =
\overline{\psi}(x)
i\sigma_{\mu\nu}T^{a}\psi(x)
\qquad {\rm and} \qquad
P^{a}_{\rm Loc}(x)
= \overline{\psi}(x) \gamma_{5}T^{a}\psi(x) \,. 
\end{equation}
The coefficients $b_{\rm V,A}$ and $c_{\rm V,A}$---together 
with $c_{\rm SW}$---can be
non-perturbatively tuned such that the cut-off effects are of
O($a^{2}$) instead of linear in $a$.\footnote{%
  Note that the mass $m$ in
  eqs.~\pref{eq:ImpV} and \pref{eq:ImpA} denotes the renormalized
  mass containing the additive mass renormalization proportional to
  $1/a$ and the renormalization factor: 
  $m = Z_{m}(m_{0} - m_{\rm cr})/a$.
  The factors of $(1+b_{\rm V,A} am)$ can also be considered to be part 
  of the renormalization factor, but it is notationally convenient 
  here to include them in the bare currents.}

In order to correctly approach the continuum limit the non-conserved
currents need to be renormalized. Thus one introduces 
\bea 
V_{\mu,{\rm ren,Loc}}^{a} = Z_{\rm V, Loc} 
\ V_{\mu,{\rm Loc}}^{a}\,, 
\qquad
&& V_{\mu,{\rm ren,Imp}}^{a} = Z_{\rm V, Imp}  
\ V_{\mu,{\rm Imp}}^{a}\,,  
\label{eq:RenV}
\\ 
A_{\mu,{\rm ren,Loc}}^{a} = Z_{\rm A, Loc} 
\ A_{\mu,{\rm Loc}}^{a}\,,  \label{eq:RenLocA} 
\qquad 
&&
A_{\mu,{\rm ren,Imp}}^{a}  =  Z_{\rm A, Imp} 
\ A_{\mu,{\rm Imp}}^{a}\,.  \label{eq:RenA} 
\eea
In the following, we will often use $Z_{\rm V,A}$ generically, without
specifying the underlying current. 

The $Z$-factors (which depend not only
on the choice of currents but also on the action)
can be fixed by imposing chiral Ward identities
\cite{Karsten:1980wd,Bochicchio:1985xa,Maiani:1986yj}. We will come
back to this important issue in section
\ref{sect:Renormalization}. The $Z$-factor for the conserved vector
current is 1, of course.

\subsection{The Symanzik effective theory}

According to Symanzik the lattice theory can be described by an
effective continuum theory provided one is close to the continuum
limit \cite{Symanzik:1983dc,Symanzik:1983gh}. This effective theory is
defined by an effective action and effective operators, and both are
strongly restricted by the locality and the symmetries of the
underlying lattice theory.  The leading terms are, by construction,
the familiar expressions of continuum QCD. Lattice artifacts appear as
higher dimensional operators multiplied by appropriate powers of the
lattice spacing.  The effective action, for example, reads
\cite{Sheikholeslami:1985ij} 
\bea
\label{eq:SymAction} S_{\rm Sym} & =
& S_{\rm ct} + a \csw \int d^{4}x\, \overline{\psi}(x) i
\sigma_{\mu\nu} F_{\mu\nu}(x)\psi(x)+ {\rm O}(a^{2}).  
\eea 
The first term is just the continuum QCD action for two flavors with degenerate
quark mass.  The leading cut-off effects are described by a single
correction term, a Pauli term containing the field strength tensor
$F_{\mu\nu}(x)$ multiplied by an unknown coefficient ('low-energy
constant') $\csw$. Many more terms appear at ${\rm O}(a^{2})$
\cite{Sheikholeslami:1985ij}.

The mapping of the bare local currents of
of eqs.~\pref{eq:LocV} and \pref{eq:LocA}
into the Symanzik theory is
~\cite{Luscher:1996sc,Sint:1997jx}: 
\bea 
V_{\mu,{\rm Loc}}^{a} &\simeq&
V_{\mu,{\rm Sym, Loc}}^{a} = 
\frac1{Z_V^0}(1+\overline{b}_{\rm V} am) \left(
V_{\mu,{\rm ct}}^{a} + a \cv \partial_{\nu} T^{a}_{\mu\nu,{\rm ct}}
\right) + {\rm O}(a^{2})\,,
\label{SymVC}
\\ 
A_{\mu,{\rm Loc}}^{a} &\simeq & 
A_{\mu,{\rm Sym, Loc}}^{a} = 
\frac1{Z_A^0}(1+\overline{b}_{\rm A} am) \left(
A_{\mu,{\rm ct}}^{a} + a \ca \partial_{\mu} P_{\rm  ct}^{a}
\right) + {\rm O}(a^{2})\,
\label{SymAC} 
\\
P_{\rm Loc}^{a} &\simeq & 
\ P_{\rm Sym, Loc}^{a}\  = 
\frac1{Z_P^0} (1+\overline{b}_{\rm P} am)P_{\rm ct}^{a} +O( a)
\label{SymP} 
\\
T_{\mu\nu,{\rm Loc}}^{a} &\simeq & 
T_{\mu\nu,{\rm Sym, Loc}}^{a} = 
\frac1{Z_T^0} (1+\overline{b}_{\rm T} am) T_{\mu\nu,{\rm ct}}^{a} +O( a)
\label{SymT} 
\eea
where the continuum bilinears take their usual forms
\be
V_{\mu,{\rm ct}}^{a} \,=\,\psibar \gamma_\mu T^{a}\psi\,,
\quad
T^{a}_{\mu\nu,{\rm ct}}\,=\, \psibar\sigma_{\mu\nu} T^{a}\psi\,,
\quad
A_{\mu,{\rm ct}}^{a} \,=\,\psibar \gamma_\mu \gamma_5 T^{a}\psi\,,
\quad
P_{\rm ct}^{a}\,=\,\psibar \gamma_{5} T^{a}\psi\ .
\ee
The mapping of operators between effective theories,
\bea
O_{\rm Lat} &\simeq & O_{\rm Sym} \,,
\eea
is defined so that 
\bea
\langle O_{\rm Lat}(x) O(\psi^{\rm Lat}, 
\bar\psi^{\rm Lat}, A_\mu^{\rm Lat}, {\bf y} ) \rangle_{S_{\rm Lat}} 
&=& \langle O_{\rm Sym}(x) O(\psi^{\rm Sym}, 
\bar\psi^{\rm Sym}, A_\mu^{\rm Sym}, {\bf y}) \rangle_{S_{\rm Sym}} \,.
\label{LatToSym}
\eea
Here $O(\psi^{\rm Lat}, \bar\psi^{\rm Lat}, A_\mu^{\rm Lat}, {\bf y} )$ 
and $O(\psi^{\rm Sym}, \bar\psi^{\rm Sym}, A_\mu^{\rm Sym}, {\bf y} )$
are arbitrary multi-local operators consisting
of quark and gluon fields at positions ${\bf y} = y_1,y_2,\cdots $
which all differ from $x$. 
The above vacuum expectation values are calculated with 
$S_{\rm Lat}$ on the left-hand-side and $S_{\rm Sym}$ on the right-hand-side. 
We note that the equations of motion have been used in order
to reduce the number of O($a$) terms on the right hand sides of
\pref{eq:SymAction}--\pref{SymAC}. This is legitimate as long as we
consider the Symanzik theory as an on-shell effective theory, set up
to reproduce physical quantities like masses, decay constants, {\em etc.}. 

We stress that the currents $V_{\mu,{\rm Sym,Loc}}^a$ and
$A_{\mu,{\rm Sym,Loc}}^a$ are the result of matching the {\em bare} lattice
currents into the Symanzik theory, 
and thus must include the renormalization 
constants $Z_{V,A}^0$. These have perturbative expansions of
the form $Z_{V,A}^0=1 + O[g(a)^2]$, but do not contain any
contributions of O($a$), since all O($a$) terms are explicitly
accounted for. They are completely determined in principle
once one has specified the action and $g(a)$, but are only
known approximately in practice. This will not present a
problem, since we ultimately will normalize the currents
non-perturbatively and all dependence on $Z_{V,A}^0$ will
cancel.
The superscript ``$0$'', which indicates that this quantity
is of 0'th order in an expansion in $a$, 
distinguishes these $Z$-factors from the
non-perturbatively determined renormalization constants $Z_{\rm V,A}$
introduced before. The latter, as we will see, depend linearly on $a$.

We can also map the renormalized currents of
eqs.~\pref{eq:RenV} and \pref{eq:RenA} into the Symanzik theory.
For $V_{\mu, {\rm ren, Loc}}$ and $A_{\mu, {\rm ren, Loc}}$,
the result will be the same as in \pref{SymVC} and \pref{SymAC}
except for multiplication by overall factors
of (the to-be-determined quantities) $Z_{\rm V, Loc}$ and 
$Z_{\rm A, Loc}$, respectively.

The results \pref{SymVC} and \pref{SymAC} also hold for
improved currents of the form of eqs.~\pref{eq:ImpV} and \pref{eq:ImpA},
although the coefficients $Z_{\rm V,A}^0$ and
$\overline c_{\rm V,A}$ will differ.
It is important to keep in mind
that these coefficients, as well as $\overline c_{\rm SW}$,
are {\em a priori} unknown,
their values depending on the details of the lattice theory. 
If all these parameters vanish 
we say that the lattice theory and current
matrix elements are O($a$)--improved. 
For this to happen the parameters $c_{\rm X}$ and
$b_{\rm X}$ of the lattice theory need to be tuned 
to appropriate nonzero values.\footnote{%
Explicitly, $c_{\rm V} = - ({Z_T^0}/{Z_V^0})\cv$, 
$c_{\rm A} = - ({Z_P^0}/{Z_A^0})\ca$, 
$b_{\rm V} = -\overline b_{\rm V}$,
and $b_{\rm A} = -\overline b_{\rm A}$.}

The vector current in \pref{SymVC} is the effective current 
into which the local lattice current is mapped. 
The conserved lattice vector current
is mapped onto the most general conserved effective vector
current. This current $V_{\mu,{\rm Sym, Con}}^{a}$ can be obtained by
starting from \pref{SymVC} (with an {\em a priori} different coefficient
$\cv$ in front of the tensor term, and with $Z_{\rm V}^0=1$,
 $\overline b_{\rm V}=0$)
and then imposing current conservation. 
The current in \pref{SymVC} is, however, already conserved, 
\bea
\label{ContdivV} 
\partial_{\mu} V_{\mu,{\rm Sym, Loc}}^{a} 
& =& 0\,,\, 
\eea 
which is a consequence of the
particular structure of the O($a$) correction (total derivative of an
antisymmetric tensor). Hence, both lattice currents are mapped onto
the same form of effective current at this order in the Symanzik expansion.
Violations of current conservation which imply a difference between
the local and the conserved current are expected to show up at
O($a^{2}$).

We mention two special cases of the Symanzik effective theory at
O($a$) that we will need in the next section for the mapping to the
chiral effective theory.  Let us first consider the Symanzik effective
theory in which $\csw$ is non-zero but $\overline c_{\rm V,A}$ and
$\overline b_{\rm V,A}$ vanish, and $Z_{V,A}^0=1$.  
In this case the Symanzik
currents coincide with the continuum QCD currents.  These transform
into each other under infinitesimal SU(2) axial rotations
$\delta\psi=\omega^aT^a\gamma_5\psi$,
$\delta\overline{\psi}=\omega^a\overline{\psi}\gamma_5T^a$:
\bea
\label{TrafoCurr} \delta V_{\mu}^a &= i \epsilon^{abc} \omega^b
A_{\mu}^c\,,\qquad \delta A_{\mu}^a &= i \epsilon^{abc} \omega^b
V_{\mu}^c\,, 
\eea 
This leads to various chiral Ward identities,
schematically written as 
$\langle \delta S_{\rm Sym} {\cal O}_{\rm  Sym}\rangle 
= \langle \delta{\cal O}_{\rm Sym}\rangle $, where
${\cal O}_{\rm Sym}$ denotes some product of vector and axial-vector
currents. The form of these Ward identities is as in continuum QCD,
since the right hand side reads 
$\langle \delta{\cal O}_{\rm Sym}\rangle 
= \langle \delta{\cal O}_{\rm ct}\rangle $. 
The only difference is the appearance of an extra term proportional to $a\csw$
in the variation of the effective action, caused by the Pauli term in
$S_{\rm Sym}$.  Note, however, that these simple QCD-like Ward
identities only hold at O($a$), and are violated at O($a^2$) by
the terms of this order in the effective action and effective currents.

The second special case is obtained by setting $\csw=0$, 
$\overline b_{\rm V,A}=0$ and $Z_{\rm V,A}^0=1$, in which case O($a$)
corrections stem entirely from the $\overline c_{\rm V,A}$
terms in the effective currents. This implies,
for example, that the correlation function of two axial-vector
currents reads 
\bea
\label{AACorr} 
\langle A_{\mu,{\rm Sym}}^a(x)
A_{\nu,{\rm Sym}}^a(y) \rangle &=& \langle A_{\mu,{\rm ct}}^a(x)
A_{\nu,{\rm ct}}^a(y) \rangle \nn\\ & & \;\; + a\ca \langle
A_{\mu,{\rm ct}}^a(x) \partial_{\nu} P_{\rm ct}^a(y) + \partial_{\mu}
P_{\rm ct}^a(x) A_{\nu,{\rm ct}}^a(y)\rangle\,.  
\eea 
Here the expectation values are defined with Boltzmann factor 
$\exp(-S_{\rm ct})$ only. 
Hence, the O($a$) correction is given by the correlation
function between the axial-vector current and the derivative of the
pseudoscalar density. Again, this result will be violated as soon as
one includes the corrections of O($a^2$).

\subsection{Matching to ChPT}
\label{ssect:MatchingtoChPT}

The appropriate chiral effective theory is obtained by writing down
the most general chiral effective Lagrangian and effective currents
which are compatible with the symmetries of the underlying Symanzik
theory. A standard spurion analysis is employed in order to properly
incorporate explicit symmetry breaking terms. For example, the
Symanzik effective action is invariant under the chiral symmetry group
$G$, parity $P$ and charge conjugation $C$, provided both the mass and
the coefficient $a\csw$ are promoted to space-time dependent external
fields $M$ and $A$, which are postulated to transform according to
\cite{Sharpe:1998xm,Rupak:2002sm,Bar:2003mh} 
\bea\label{trafoMandA}
\ba{ccccc} M \stackrel{G}{\longrightarrow} LMR^{\dagger},&\quad& M
\,\stackrel{P}{\longrightarrow} M^{\dagger},\,&\quad& M
\,\stackrel{C}{\longrightarrow} M^{T}\,,\\ 
A \stackrel{G}{\longrightarrow} LAR^{\dagger},&\quad& A
\,\stackrel{P}{\longrightarrow} A^{\dagger},\,&\quad& A
\,\stackrel{C}{\longrightarrow} A^{T}\,.  \ea 
\eea 
The ``physical''
values are obtained by setting $M\rightarrow m$ and $A\rightarrow
a\csw$. In an intermediate step, however, the spurion fields $M$ and
$A$ are used together with the standard field 
\bea 
\Sigma(x) & =&
\exp\left(\frac{2i}{f}\pi^{a}(x)T^{a}\right) 
\eea 
in order to write
down the most general scalar that is invariant under $G,\, P$ and
$C$. This has been done in Refs.\ \cite{Rupak:2002sm,Bar:2003mh}, and
part of the result reads (in Euclidean space-time) 
\bea 
{\cal L}_{\rm
  chiral} & = & \frac{{f^2 }}{4}\left\langle \partial_{\mu} \Sigma
(\partial_{\mu} \Sigma)^\dagger \right\rangle - \frac{{f^2}}{4}
2Bm\Tr{\Sigma^{\dagger} + \Sigma}\nn\\ & &
+L_{45}2Bm\Tr{\partial_{\mu} \Sigma (\partial_{\mu} \Sigma)^\dagger}
\Tr{\Sigma^{\dagger} + \Sigma\nn}
\nn\\ &&+W_{45}\hat{a}\csw\Tr{\partial_{\mu} \Sigma (\partial_{\mu}
  \Sigma)^\dagger} \Tr{\Sigma^{\dagger} + \Sigma} \nn \\
  & & - W_{68} 2Bm\, \hat{a}\csw \langle \Sigma +\Sigma^{\dagger} \rangle^2 + 
\ldots\label{Lag}\,.  
\eea 
Here the angled brackets denote traces in
flavor space.\footnote{%
  In the last section we used the same notation
  for correlation functions. The context usually tells unambiguously
  what is meant  by $\langle \ldots\rangle$.}
The lattice spacing appears in the combination 
\bea
\hat{a} & = & 2W_{0}a\,, 
\eea 
which is of dimension
two.\footnote{%
$W_{0}$, a LEC that enters the chiral Lagrangian at
  O($a$), is of dimension three \cite{Rupak:2002sm}.} 
We have dropped a number
of terms of O($p^{4}$) and O($a^{2}$), which we will not need
in the following. Note that we have absorbed the term $f^2 W_{0}a\csw
\Tr{\Sigma^{\dagger} + \Sigma}/2$ in the definition of the mass, so
$m$ in \pref{Lag} corresponds already to the so-called shifted mass
\cite{Sharpe:1998xm,Sharpe:2004ny}.

Taking the naive continuum limit $a\rightarrow0$ we recover the
correct terms of the continuum chiral Lagrangian with the familiar
low-energy coefficients $f$, $B$ and $L_{45} = L_{4} + L_{5}/2$ of
continuum ChPT \cite{Gasser:1983yg,Gasser:1984gg}. $W_{0}$ and $W_{45}
= W_{4} + W_{5}/2$ are low-energy constants associated with breaking
terms due to the nonzero lattice spacing
\cite{Rupak:2002sm}.\footnote{%
  Our notation for the low-energy
  coefficients differs slightly compared to the notation in other
  references, since our $m$ is already the shifted mass: our $W_{45}\csw$
  and $W_{68}\csw$ 
  correspond to the combinations $W_{45} - L_{45}$ and $W_{68} - 2L_{68}$ in
  Ref.\ \cite{Rupak:2002sm}. These combinations are abbreviated to
  $\widetilde{W}$ and $W$ in Ref.\ \cite{Sharpe:2004ny}.}

We now apply the same procedure to derive expressions for the
effective operators.  The currents in the Symanzik effective theory
are given in eqs.~\pref{SymVC} and \pref{SymAC}---forms
which, as noted above, hold for all the lattice currents of interest.
To simplify the following discussion, we will map the
Symanzik currents {\em without} overall $Z$-factors and $b_{\rm X}$ 
corrections into the chiral effective theory. These factors
can be added 
at the end. Thus we consider the mappings
\bea
V_{\mu,{\rm ct}}^{a} + a \cv \partial_{\nu} T^{a}_{\mu\nu,{\rm ct}}
&\longrightarrow& V_{\mu,{\rm eff}}^a \,,
\\
A_{\mu,{\rm ct}}^{a} + a \ca \partial_{\mu} P_{\rm  ct}^{a}
&\longrightarrow& A_{\mu,{\rm eff}}^a \,.
\eea
The leading
contributions are just the continuum expressions for the currents,
\bea 
V_{\mu,{\rm ct}}^{a} & = & 
\psibar_{R} \gamma_{\mu} T^{a}\psi_{R}
+ \psibar_{L} \gamma_{\mu} T^{a}\psi_{L}\,, 
\\ 
A_{\mu,{\rm ct}}^{a} &= & 
\psibar_{R} \gamma_{\mu} T^{a}\psi_{R} - \psibar_{L} \gamma_{\mu}
T^{a}\psi_{L}, 
\eea 
where we decomposed the currents into chiral
fields. The vector current is just the sum of the right and left
handed current, the axial-vector current is given by the difference.

The O($a$) corrections couple fields with opposite chirality, and the
currents in the Symanzik theory do not transform under chiral
rotations as continuum vector and axial-vector currents. However, the
continuum transformation behavior can be enforced by promoting the
coefficients $\cv,\ca$ to spurion fields $C_{V}$ and $C_{A}$ with
non-trivial transformation behavior under $G$, $P$ and $C$:
\bea\label{TrafoCs} C_{X} & \stackrel{G}{\longrightarrow} &
LC_{X}R^{\dagger}\,,\qquad C_{X} \, \stackrel{P} {\longrightarrow} \,
C_{X}^{\dagger} ,\qquad C_{X}\, \stackrel{C}{\longrightarrow}\,
C_{X}^{T}\,,\qquad X\,=\,V,A\,.  \eea Note that these transformation
laws are identical to the ones of the other O($a$) spurion $A$,
cf.\ \pref{trafoMandA}.

It is now easily checked that the O($a$) corrections written in the
form 
\bea 
V_{\mu, a{\rm-corr}}^{a} & = & (\partial_{\nu}\psibar_{L})
i\sigma_{\mu\nu} C_{V} T^{a}\psi_{R} + \psibar_{L} i\sigma_{\mu\nu}
T^{a} C_{V}\partial_{\nu}\psi_{R} \nonumber
\\ 
& & +(\partial_{\nu}\psibar_{R}) i\sigma_{\mu\nu} C_{V}^{\dagger}
T^{a}\psi_{L} + \psibar_{R} i\sigma_{\mu\nu} T^{a}
C_{V}^{\dagger}\partial_{\nu}\psi_{L},\label{Vmuacorr}
\\ 
A_{\mu,a{\rm-corr}}^{a}
& = & (\partial_{\mu}\psibar_{L})\gamma_{5} C_{A} T^{a}\psi_{R} +
\psibar_{L} \gamma_{5}T^{a} C_{A}\partial_{\mu}\psi_{R} \nonumber
\\ 
& & +(\partial_{\mu}\psibar_{R})\gamma_{5} C_{A}^{\dagger} T^{a}\psi_{L}
+ \psibar_{R} \gamma_{5}T^{a}
C_{A}^{\dagger}\partial_{\mu}\psi_{L},\label{Amuacorr} 
\eea 
transform as the continuum currents. Setting the spurion field to its
``physical'' value, $C_{V} \rightarrow a\cv$ and $C_{A} \rightarrow
a\ca$, one recovers the desired Symanzik currents.

The currents in the chiral effective theory are now obtained by
writing down the most general vector and axial-vector current, build
of the chiral field $\Sigma$, its derivatives and the spurion fields.
The spurions necessary for the construction of the vector current are
$M$ and $A$, the ones we already encountered in the construction of
the effective action, and $C_{V}$.  For the axial-vector $C_{A}$ must
be used instead.

In order to write down the currents we first recall that our effective
theory has to reproduce continuum ChPT if we send
$a\rightarrow0$. This requirement implies that the continuum part of
the currents are just the expressions given by Gasser and Leutwyler.
At leading order these read 
\bea 
V_{\mu,{\rm LO}}^{a}
&=&\frac{f^{2}}{2}\Tr{T^{a}(\Sigma^{\dagger}\partial_{\mu}\Sigma +
  \Sigma\partial_{\mu}\Sigma^{\dagger} )}\,,
\\ 
A_{\mu,{\rm LO}}^{a}
&=& \frac{f^{2}}{2}\Tr{T^{a}(\Sigma^{\dagger}\partial_{\mu}\Sigma -
  \Sigma\partial_{\mu}\Sigma^{\dagger} )}\,.  
\eea 
Obviously these
expressions transform as vector and axial-vector currents under $G$,
$P$ and $C$. Moreover, these currents are properly normalized in order
to satisfy the current algebra.

In order to construct the leading O($a$) corrections we need at least
one power of either $A$ or $C_{X}$ and one partial derivative of
$\Sigma$. It is easily checked that the following terms transform as
desired: 
\bea 
\ba{rcl} 
V1) && V_{\mu,{\rm LO}}^{a}
\Tr{\Sigma^{\dagger}A + \Sigma A^{\dagger}},
\\[0.2ex] 
V2) &&
V_{\mu,{\rm LO}}^{a} \Tr{\Sigma^{\dagger}C_{V} + \Sigma
  C_{V}^{\dagger}},
\\[0.2ex] 
V3) && \Tr{T^{a}(\partial_{\mu}
  \Sigma^{\dagger} A - A \partial_{\mu} \Sigma^{\dagger} +
  \partial_{\mu} \Sigma A^{\dagger} - A^{\dagger} \partial_{\mu}
  \Sigma)}, 
\\[0.2ex] 
V4) && \Tr{T^{a}(\partial_{\mu} \Sigma^{\dagger}
  C_{V} - C_{V} \partial_{\mu} \Sigma^{\dagger} + \partial_{\mu}
  \Sigma C_{V}^{\dagger} - C_{V}^{\dagger} \partial_{\mu}
  \Sigma)}, \label{V1} 
\ea 
\eea 
for the vector current, and
\bea
\label{A2} \ba{rcl} 
A1) &&A_{\mu,{\rm LO}}^{a}
\Tr{\Sigma^{\dagger}A + \Sigma A^{\dagger}},
\\[0.2ex] 
A2) &&A_{\mu,{\rm LO}}^{a} \Tr{\Sigma^{\dagger}C_{A} + \Sigma
  C_{A}^{\dagger}},
\\[0.2ex] 
A3) && \Tr{T^{a}(\partial_{\mu}
  \Sigma^{\dagger} A + A \partial_{\mu} \Sigma^{\dagger} -
  \partial_{\mu} \Sigma A^{\dagger} + A^{\dagger} \partial_{\mu}
  \Sigma)},
\\[0.2ex] 
A4) && \Tr{T^{a}(\partial_{\mu} \Sigma^{\dagger}
  C_{A} + C_{A} \partial_{\mu} \Sigma^{\dagger} - \partial_{\mu}
  \Sigma C_{A}^{\dagger} + C_{A}^{\dagger} \partial_{\mu} \Sigma)},
\ea 
\eea 
for the axial-vector current.  Setting the external fields to
their final value, $C_{X}\rightarrow a \overline{c}_{X}$, we obtain the
following expressions for the currents in the effective theory: 
\bea
V_{\mu}^{a} & = &V_{\mu,{\rm LO}}^{a}\left(1 +
\frac{4}{f^{2}}\hat{a}\big({W}_{V1}\csw+{W}_{V2}\cv\big) \Tr{\Sigma +
  \Sigma^{\dagger}}\right) ,\label{Vdirect}
\\ 
A_{\mu}^{a} & = &
A_{\mu,{\rm LO}}^{a}\left(1 +
\frac{4}{f^{2}}\hat{a}\big({W}_{A1}\csw+{W}_{A2}\ca\big) \Tr{\Sigma +
  \Sigma^{\dagger}} \right)\nn
\\[0.2ex] 
& & \;\; +
4\hat{a}\big({W}_{A3}\csw+{W}_{A4}\ca\big)\partial_{\mu}
\Tr{T^{a}(\Sigma - \Sigma^{\dagger})} \label{Adirect}\,.  
\eea 
The coefficients ${W}_{X}$ are unknown low-energy constants (LECs). 
In order to make ${W}_{X1},{W}_{X2}$ dimensionless we included the factor
$4/f^{2}$.
Note that the
second line in \pref{Adirect} is proportional to the continuum 
pseudoscalar density, 
$P_{\rm ct}^a=f^2B\Tr{T^{a}(\Sigma -  \Sigma^{\dagger})}/2$.

The number of unknown LECs in the currents can be reduced using the
freedom of field redefinition
\cite{Sharpe:2004ny,Scherer:1994wi}. 
Explicitly, performing
$\Sigma\rightarrow \Sigma +\delta \Sigma$ with 
\bea 
\delta \Sigma &=&
\frac{4\hat{a}}{f^2}\csw \Delta W \big(\Sigma^2 - 1\big)\,, 
\eea 
we obtain the same effective Lagrangian and currents with the transformed
coefficients 
\bea 
\ba{rclcrcl} 
W_{45}&\rightarrow& W_{45} + \Delta
W\,,& \quad & W_{68} & \rightarrow & W_{68} + \Delta W/2\,,\\ 
W_{V1}&\rightarrow& W_{V1} + \Delta W\,,& \quad &  &
 &\\
 W_{A1} & \rightarrow & W_{A1} + \Delta W\,,& \quad & W_{A3} &
\rightarrow & W_{A3} - \Delta W\,.
\ea 
\eea 
Therefore, depending on
the particular choice for $\Delta W$ we may make one LEC vanish. In
the following we choose this to be $W_{A3}$ in the expression for the
axial-vector current.

So far the expressions in \pref{Vdirect} and \pref{Adirect} are the
most general currents compatible with the symmetries that have the
correct continuum limit. In order to match the currents properly we
have to impose the constraints that these currents have to obey at
O($a$), for instance current conservation for the vector current. This
will relate some of the LECs ${W}_{X}$ to those in the effective
action.

In order to discuss this we first derive the EOM corresponding to the
Lagrangian in eq.\ \pref{Lag} without the continuum NLO term
proportional to $L_{45}$ and without the O($am$) correction proportional to $W_{68}$.\footnote{%
  These can and need to be included if
  the NLO terms of O$(p^{2}m$) and O$(p^{2}ma$) are included in the current.}
Using the shorthand notation $\chi = 2Bm$, $\rho = 2W_{45} W_0a\csw$ and
\bea 
R & =& \frac{4}{f^2}\Tr{\Sigma+\Sigma^{\dagger}},\qquad T \,=\,
\frac{4}{f^2}\langle \partial_{\mu}
\Sigma\partial_{\mu}\Sigma^{\dagger}\rangle\,, 
\eea 
the leading order EOM reads 
\bea 
& \Big[\Sigma (\partial_{\mu} \partial_{\mu}
  \Sigma^{\dagger}) - (\partial_{\mu} \partial_{\mu} \Sigma
  )\Sigma^{\dagger} \Big] (1+\rho R) -\Big[\Sigma
  -\Sigma^{\dagger}\Big] (\chi + \rho T )& \nn
\\ 
& = 2\lambda + \Big[
  \partial_{\mu}\Sigma \Sigma^{\dagger} - \Sigma
  \partial_{\mu}\Sigma^{\dagger}\Big] \rho \partial_{\mu}R
\,. \label{EOMWChPT} 
\eea 
The parameter $\lambda$ is the Lagrange
multiplier associated with the constraint $\det \Sigma =1$. Setting
the lattice spacing to zero (i.e.\ $\rho=0$) eq.\ \pref{EOMWChPT}
reproduces the EOM in continuum ChPT \cite{Gasser:1984gg}.

Using \pref{EOMWChPT} we find the condition 
\bea
\label{CondVCCons}
\partial_{\mu} V_{\mu}^{a} & = & 0 \quad \Leftrightarrow \quad W_{V1}
= W_{45}\,\,\,{\rm and} \,\,\, W_{V2} = 0\,.  
\eea 
Therefore, the
conserved vector current in the chiral effective theory is given by
\bea 
V_{\mu, {\rm eff}}^{a} & = &V_{\mu,{\rm LO}}^{a}\left(1 +
\frac{4}{f^{2}}\hat{a}{W}_{45} \csw \Tr{\Sigma +
  \Sigma^{\dagger}}\right)\,.
\label{VeffNoether} 
\eea 
This expression
agrees with the one in Ref.\ \cite{Sharpe:2004ny} obtained from the
generating functional.  Note that at this order this current coincides
with the Noether current associated with vector transformations. 
This will, however, no longer be true at higher order 
in the chiral expansion.\footnote{%
  At higher order in the chiral
  expansion the currents receive contributions which are not present
  in the Noether current. The vector current contribution proportional
  to $L_9$ \cite{Gasser:1984gg}, which captures the dominant
  contribution of the pion form factor, is an example for this. We
  thank J.\ Bijnens for pointing this out to us.}

Note that there is no term proportional to $\cv$ in
\pref{VeffNoether}, which is a consequence of the 
$\partial_{\nu} T^{a}_{\mu\nu}$ 
structure of the O($a$) correction. We need three
partial derivatives in order to construct such a term in the chiral
effective theory. Hence, this correction is of O($ap^3$), which is
of higher order than we consider here.

In order to obtain the proper result for the axial-vector current we
have to make sure that the properties \pref{TrafoCurr} and
\pref{AACorr} of the underlying theory are correctly reproduced. The
result \pref{VeffNoether} for the vector current together with
\pref{TrafoCurr} (for $\cv=\ca=0$) immediately implies
$W_{A1}=W_{45}$. On the other hand, setting $\csw=0$ and demanding
\pref{AACorr} leads to $W_{A2}=0$. So the final result for the axial
vector current reads 
\bea 
A_{\mu,{\rm eff}}^{a} & = & A_{\mu,{\rm
    LO}}^{a}\left(1 + \frac{4}{f^{2}}\hat{a}{W}_{45}\csw \Tr{\Sigma +
  \Sigma^{\dagger}} \right) + 4\hat{a}{W}_{A}\ca\partial_{\mu}
\Tr{T^{a}(\Sigma - \Sigma^{\dagger})} 
\label{Adirect2}\,, 
\eea 
where we abbreviated $W_{A4}=W_A$.  This expression also agrees with the
result in Ref.~\cite{Sharpe:2004ny}.\footnote{%
  The result in
  Ref.\ \cite{Sharpe:2004ny} is obtained by setting $W_{10}=2W_A\ca$
  and $\widetilde W=2 W_{45} \csw$.}

So far we have discussed the currents at leading order in the chiral
expansion including the first correction of O($a$). Higher order
contributions to the currents can be derived in the same fashion. The
terms without factors of the lattice spacing are the familiar
contributions of continuum ChPT, 
\bea 
\delta V_{\mu,{\rm eff}}^{a} & =
&\frac{f^{2}}{2} \Tr{T^{a}(\Sigma^{\dagger}\partial_{\mu}\Sigma +
  \Sigma\partial_{\mu}\Sigma^{\dagger})}\left(\frac{4}{f^{2}}2BmL_{45}
\Tr{\Sigma + \Sigma^{\dagger}}\right) ,\label{dVNLO}
\\ 
\delta A_{\mu,{\rm eff}}^{a} & = & \frac{f^{2}}{2}
\Tr{T^{a}(\Sigma^{\dagger}\partial_{\mu}\Sigma -
  \Sigma\partial_{\mu}\Sigma^{\dagger})}\left(\frac{4}{f^{2}}2Bm
L_{45}\Tr{\Sigma + \Sigma^{\dagger}} \right). \label{dANLO} 
\eea 
In order to construct the first subleading O($a$) corrections we have to
form vector and axial-vector currents with one power of either $A$ or
$C_{X}$, and either three derivatives (corresponding to the O($ap^{2}$
contributions), or one derivative and one power of the mass spurion
field $M$ (the O($am$) contributions).  Simple examples for such terms
are products of the leading O($a$) terms in \pref{V1}--\pref{A2} with
either $\Tr{\Sigma M^{\dagger} + M\Sigma^{\dagger}}$ or
$\Tr{\partial_{\mu} \Sigma (\partial_{\mu} \Sigma)^\dagger}$. An 
example is the correction 
\bea 
\delta A_{\mu,{\rm eff}}^{a} & = &
\frac{f^{2}}{2} \Tr{T^{a}(\Sigma^{\dagger}\partial_{\mu}\Sigma -
  \Sigma\partial_{\mu}\Sigma^{\dagger})}\left(X \ca \hat{a}m\Tr{\Sigma
  + \Sigma^{\dagger}}^{2} \right), \label{dANLO2} 
\eea 
with $X$ being the low-energy constant associated with this correction.

There are more terms possible and it is a straightforward but tedious
exercise to find all possible terms contributing to the currents. For
the rest of this paper, however, we do not need these higher-order
terms.

We conclude this section by discussing the pseudoscalar density, which we need
later on for the computation of the PCAC mass.  In the Symanzik
effective theory the pseudoscalar density is given by 
$P^a_{\rm ct}$.
The corresponding expression in the chiral effective theory is
constructed in the same way as the currents. Since the density in the
Symanzik theory has no O($a$) term, we only have the spurion $A$ for
the construction of the density, and we find\footnote{%
  This agrees with the result of Ref.\ \cite{Sharpe:2004ny}, 
  with $W_{68}$ related to $W$ of that reference by $W=W_{68}\csw$.}
\bea
\label{eq:Pseudoscalareff} 
P^{a}_{\rm eff} & = & \frac{f^{2}B}{2}
\Tr{T^{a}(\Sigma^{\dagger} - \Sigma)}\left(1 +
\frac{4}{f^{2}}\hat{a}{W}_{68}\csw \Tr{\Sigma +
  \Sigma^{\dagger}}\right)\,.
\eea 

As already mentioned, our expressions for the effective currents
agree with the results in Ref.~\cite{Sharpe:2004ny}, although the
two calculations use different methods.
In the approach of Ref.~\cite{Sharpe:2004ny},
which follows the procedure
used by Gasser and Leutwyler in continuum ChPT
\cite{Gasser:1983yg},
sources for the currents are introduced and the
generating functional is constructed.
The currents are then obtained from the
generating functional by taking derivatives with respect to the
sources. The analysis is complicated,
however, by the presence in the Symanzik theory
of O($a$) violations of the local gauge invariance used to  
restrict the mapping to the chiral effective theory.
This complication, missed in Ref.~\cite{Sharpe:2004ny},
was noted in Ref.~\cite{SharpeNara}, and an appropriate extension outlined.
The outcome of this extension was that the form of the results
of Ref.~\cite{Sharpe:2004ny} still hold after a redefinition of 
the LECs. Since a systematic and complete treatment of the impact
of the O($a$) effects in the generating
functional method has not been presented, however, we
provide such a treatment in appendix \ref{sourcemethod}.
In particular, we show that the same results 
\pref{VeffNoether} and \pref{Adirect2} are obtained 
using this method.

\section{Renormalization}
\label{sect:Renormalization}

We now return to the issue of the normalization of the currents.
The results of the previous section allow us to determine
the form that a given lattice current will have when mapped
into WChPT. For example, putting back the overall factors, 
the renormalized local vector and axial-vector currents 
of eqs.~\pref{eq:RenV} and \pref{eq:RenA} map as
\begin{eqnarray}
V_{\mu,\rm ren,Loc} &\simeq& 
\frac{Z_{\rm V,Loc}}{Z_{\rm V}^0}(1 + \overline b_{\rm V} am)
V_{\mu,\rm eff}^a\,,\label{eq:VlocWChPT}
\\
A_{\mu,\rm ren,Loc} &\simeq& 
\frac{Z_{\rm A,Loc}}{Z_{\rm A}^0}(1 + \overline b_{\rm A} am) 
A_{\mu,\rm eff}^a\,,\label{eq:AlocWChPT}
\end{eqnarray}
with $V_{\mu,\rm eff}^a$ and
$A_{\mu,\rm eff}^a$ given by eqs.~(\ref{VeffNoether})
and (\ref{Adirect2}), respectively.
In the lattice theory, the renormalization factors are determined
by imposing particular conditions at nonzero lattice spacing. 
Hence, in order to properly match the
effective currents we should impose the same conditions in the
effective theory.

\subsection{Renormalization of the lattice currents}
Since the conserved vector current does not need to be renormalized it
can be used to normalize the local current and to define $Z_{\rm V,Loc}$
(at the lattice level)
according to \cite{Martinelli:1990ny,Martinelli:1993dq}
\bea
\label{eq:ZVLatt1} 
Z_{\rm V,Loc} & = & \frac{\langle f | \VC{a}{Con} |
  i\rangle}{\langle f | \VC{a}{Loc} | i\rangle}.  
\eea 
Here $i$ and
$f$ denote arbitrary initial and final states, though it is
convenient to choose zero-momentum pseudoscalar states. Note that the
dependence of $Z_{\rm V,Loc}$ on the particular states can be sizable, in
particular if the theory is not O($a$)--improved
\cite{Martinelli:1990ny}.

An alternative definition for $Z_{\rm V,Loc}$, which does not
use the conserved current, is given by
\cite{Martinelli:1990ny,Martinelli:1993dq,Henty:1994ta}
\bea
\label{eq:ZVLatt2} 
Z_{\rm V,Loc} \langle \pi^{a}(\vec{p}) 
| V^{b}_{0,{\rm Loc}} | \pi^{c}(\vec{p}) \rangle& = & \epsilon^{abc} 2 E\,.  
\eea
The matrix element on the left-hand-side can be obtained in the usual
way by calculating the ratio of two correlation functions, where
pseudoscalar sources are used to project onto the pion
states. Although we have specified initial and final 
pion states, the renormalization factor still depends on the momenta
of the pions \cite{Martinelli:1990ny}. For massive pions one usually
chooses zero spatial momentum, $\vec{p}=0$.

The two renormalization conditions \pref{eq:ZVLatt1} and
\pref{eq:ZVLatt2} do not specify the renormalization condition
completely, since the matrix elements still depend on the quark
mass. Theoretically preferable are mass independent renormalization
schemes, in which the renormalization condition is imposed at zero quark
mass.  This implies some technical difficulties, because numerical
simulations cannot be performed directly for vanishing quark
masses. One way to circumvent this technical limitation is to
calculate the $Z$-factors for various small quark masses and
extrapolate the results to the massless limit. This may introduce some
extrapolation error, but in principle is a viable procedure.

A different strategy to define and compute the renormalization factors
makes use of Schr\"odinger functional boundary conditions
\cite{Jansen:1995ck}. With this setup it is possible to compute
$Z_{\rm V,Loc}$ and other renormalization factors for a vanishing partially
conserved axial-vector current (PCAC) quark mass. Even though
the renormalization conditions are imposed at vanishing quark mass,
they now depend on details of the Schr\"odinger functional setup,
e.g. the size and geometry of the finite volume. This
dependence can, in principle, be removed by extrapolating to infinite
volume.

Being able to work at vanishing quark mass has another advantage:
chiral Ward identities involving the axial-vector current simplify
significantly. In Ref.~\cite{Luscher:1996jn},
for example, the Ward identity 
\bea
\label{eq:axialvectorWI}
\int_{\partial R} {\rm d}\sigma_{\mu}(x) \epsilon^{abc} \langle f|
A_{\mu, {\rm ren}}^{a}(x) A_{\nu,{\rm ren}}^{b}(y) |i\rangle = 2i
\langle f | V_{\nu,{\rm ren}}^{c}(y) | i \rangle 
\eea 
is imposed.  This
identity is the Euclidean analogue of the current algebra relation
stating that the commutator of two axial-vector currents is equal to
the vector current \cite{Luscher:1998pe}.  The region $R$ is chosen
to be the space-time volume between two hyper-planes at $x_{0}=y_{0}
\pm t$.  The equation for $\nu=0$ between pseudoscalar states has
been used to determine the renormalization factor $Z_{A}$. For more
details see Ref.~\cite{Luscher:1996jn}.

\subsection{Renormalization of the effective currents}
Having chosen particular renormalization conditions for the lattice
currents, we have to impose the same conditions in the chiral
effective theory. 

Some conditions are harder to implement than others. For example,
matrix elements between the vacuum and a vector meson state in
eq.~\pref{eq:ZVLatt1} are not easily accessible in standard mesonic
ChPT, since the vector meson is not a degree of freedom in the chiral
effective theory.  Conditions involving quark states (the so-called
``RIMOM'' scheme \cite{Martinelli:1994ty}) are also out of reach.  In
practice, only conditions involving pseudoscalar states can be
treated in the chiral effective theory.

In the following we carry out the matching for one particular class
of renormalization conditions. For the vector current we assume that
either the condition \pref{eq:ZVLatt1} or \pref{eq:ZVLatt2} is imposed
with single pion states at zero spatial momentum and at vanishing bare
PCAC mass. For the axial-vector current we assume that condition
\pref{eq:axialvectorWI} is employed to fix $Z_{A}$. We impose these
conditions in infinite volume. Finite volume can also be considered,
but it does not make a difference at the order in the chiral expansion
to which we work.

As a first step we have to calculate the PCAC mass and set it to zero.
The PCAC mass is defined by 
\bea 
m_{\rm PCAC} & =& \frac{\langle 0 |
  \partial_{\mu} A^{a}_{\mu,{\rm eff}}| \pi^{a}\rangle}{2\langle 0 |
  P^{a}_{\rm eff}| \pi^{a}\rangle}\,.  
\eea 
Expanding the $\Sigma$
fields in \pref{Adirect2} and \pref{eq:Pseudoscalareff}, keeping only
one power of $\pi^{a}$, the ratio of expectation values on the right
hand side is easily obtained\footnote{%
In practice, to calculate $m_{\rm PCAC}$ requires knowledge
of the renormalization constants of the lattice axial-vector current
and pseudoscalar density, which can, as the present work shows,
introduce additional O($a$) corrections. These do not, however,
change the key result being derived here, namely that 
$m_{\rm PCAC}\propto m$ up to O($a^2$) corrections.}
\bea
\label{PCACmass} 
m_{\rm PCAC} & =&
\frac{M_{\pi}^{2}}{2B}\left(1 +
\frac{8}{f^{2}}\hat{a}[2(W_{45}-W_{68})\csw + W_{A}\ca]\right),
\eea
in agreement with Ref.~\cite{Sharpe:2004ny}.
The PCAC quark mass is proportional to the pion mass, 
which is given by~\cite{Sharpe:2004ny}
\bea
\label{eq:Mpitree} 
M_{\pi}^{2} & = & 2Bm\left(1 +
\frac{16}{f^2} \hat{a} (2 W_{68}-W_{45})\csw\right).  
\eea 
Recall that $m$ denotes
the shifted mass including the leading O($a$) shift.

At higher order in the chiral expansion the right hand side of
\pref{eq:Mpitree} contains an additional correction of O($a^{2})$
\cite{Sharpe:1998xm}. A vanishing pion mass therefore corresponds to
$m={\rm O}(a^{2})$.  Since we ignore ${\rm O}(a^{2})$ corrections, we 
conclude that a vanishing PCAC quark mass is equivalent to $m=0$, and
in the following we assume this condition for the shifted
mass.\footnote{%
  The O($a^{2})$ correction implies a non-trivial phase
  structure of the theory with two qualitatively different scenarios
  \cite{Sharpe:1998xm}. One of these is characterized by a first order
  phase transition and the pion mass is non-zero for all values of
  $m$. For $m=0$, however, the pion mass assumes its minimal value
  $M^2_{\pi,{\rm min}}={\rm O}(a^{2})$. Since here we ignore the
  O($a^2$) corrections we also find for this scenario that a vanishing
  PCAC mass is given by a vanishing shifted mass, at least to the
  order we are working.}

The next step is the determination of $Z_{\rm V,Loc}$
using either \pref{eq:ZVLatt1} or \pref{eq:ZVLatt2}.
Both conditions are easily calculated
using the chiral effective theory ``image'' of the local lattice
current, eq.~\pref{eq:VlocWChPT} with $m=0$, and the expression \pref{VeffNoether}.
In both cases we find\footnote{%
It is a simple matter to restore the mass dependence, in which case
\pref{eq:ZV} would read 
$Z_{\rm V,Loc}  =  Z_{\rm V}^0(1+\overline{b}_{\rm V} am)$.
}
\bea
\label{eq:ZV} 
Z_{\rm V,Loc} & = & Z_{\rm V}^0 
\,.  
\eea
Even though the result is the same for both renormalization
conditions, the way it arises differs in the two cases.
The result for the first condition is obviously
trivial since both the local and the conserved effective vector
currents have the same form in the effective theory
[cf.\ \pref{VeffNoether}], differing only by overall factors.  
In \pref{eq:ZVLatt2}, on the other hand, the two pion states contribute a
wave function renormalization factor $Z_{\pi}$, which reads\footnote{%
Note that there are no terms proportional to the quark
mass in the chiral Lagrangian \pref{Lag}, since we have set $m$ to zero.}
\bea
\label{eq:Zpi} 
Z_{\pi}& = & 1 - \frac{16}{f^{2}} \hat{a} W_{45}\csw, 
\eea 
and which exactly cancels the O($a$) correction coming
from the current.

Let us consider higher order corrections to the result
\pref{eq:ZV}. Expanding the $\Sigma$ fields in the vector current
\pref{VeffNoether} the first correction terms contain four pion
fields. Two of these need to be contracted and form a loop. Hence the
result will be proportional to $M_{\pi}^{2}\ln M_{\pi}^{2}$, which
vanishes for $m_{\rm PCAC}=0$.  Higher order analytic parts, on the
other hand, will contribute corrections of O($a^{2}$), which we do not
consider here.

The implication of eq.~\pref{eq:ZV} is that, after one or other
of the renormalization conditions has been enforced, the
renormalized local vector current maps simply into $V_{\mu,\rm eff}^a$,
since the prefactors on the right-hand-side of eq.~\pref{eq:VlocWChPT}
cancel. The same holds true for any lattice vector current which
is renormalized in this way.

We now proceed to the local axial-vector current, whose normalization
$Z_{\rm A,Loc}$ is to be fixed by imposing
Ward identity \pref{eq:axialvectorWI}, using
external one-pion states having definite (nonzero) momenta.
Following
Ref.~\cite{Luscher:1996jn} we choose the region $R$ to be the
space-time volume between two hyper-planes at $x_{0}=y_{0} \pm t$ with
some finite time separation $t$. The equation for $\nu=0$ can be
brought into the form \cite{Luscher:1996jn} 
\bea
\label{AVWI3} 
\int {\rm d}\vec{x}\, \epsilon^{abc}\epsilon^{cde} \langle
\pi^{d}(\vec{p})| [A_{0,{\rm ren}}^{a}(y_{0}+t,\vec{x}) - 
  A_{0,{\rm ren}}^{a}(y_{0}-t,\vec{x})] A_{0,{\rm ren}}^{b}(y) |
\pi^{e}(\vec{q})\rangle \nonumber\\ 
= 2i \epsilon^{cde} \langle
\pi^{d}(\vec{p})| V_{0,{\rm ren}}^{c}(y) |\pi^{e}(\vec{q}) \rangle\,,
\eea 
The matrix element on the right hand side of this equation is
essentially the one in \pref{eq:ZVLatt2} that we used to 
fix $Z_{\rm V,Loc}$.
Imposing \pref{AVWI3} thus provides a condition for $Z_{\rm A,Loc}$:
simply compute the two sides in the effective theory and
set them equal.
The calculation is straightforward but rather technical. For
this reason we defer the details of the calculation to appendix
\ref{app_WTI}. 
The final result is (recall that $m=0$ so there is no $\overline{b}_{\rm A}$
term)
\bea
\label{ZAWChPT}
\frac{Z_{\rm A,Loc}}{Z_A^0} 
&=&
1 - \frac{4\hat{a}}{f^2}(W_{45}\csw+W_A\ca) z_A(t)\,,\label{ZAresult}
\\
z_A(t) &=&
1 - \cosh[t(|\vec p|-|\vec q|)] \exp[-|t| |\vec p - \vec q|]\,.
\label{eq:zAt}
\eea 
Since this result is determined by a ratio of physical
correlation functions in which the axial currents are separated,
it must depend on the ``physical'' combination of LECs
$W_{45}\csw+W_A\ca$ \cite{Sharpe:2004ny}. 
This point is also explained
at the end of appendix \ref{sourcemethod}, 
and provides a check of our result.
We emphasize that \pref{ZAresult} is the complete result to O($a$), there are no corrections of O($am$) since we work at zero quark mass. The next correction to \pref{ZAresult} is of O($a^2$), which we do not consider here.

In contrast to the vector current we do find a non-vanishing correction
of O($a$).  
That this correction depends on the separation between the axial
currents, $t$, and upon the external state, is as expected.
The $t$ dependence is proportional to an integral
(a sum on the lattice) of the divergence of the
axial current, which does not vanish at O($a$), even when $m=0$.
The dependence of the external state is a generic feature of
O($a$) corrections in an unimproved theory.

Using eqs.~\pref{eq:AlocWChPT} and \pref{ZAWChPT} 
we find that the renormalized local axial-vector
current maps into WChPT as
\begin{eqnarray}
A_{\mu,\rm ren,Loc}^a &\simeq& 
A_{\mu,\rm ren,Loc,eff}^a =
\left[1 - \frac{4\hat{a}}{f^2}(W_{45}\csw+W_A\ca) z_A(t)\right]
A_{\mu,\rm eff}^a\,.
\label{eq:AlocWChPT_final}
\end{eqnarray}
This is the main new result of this paper.
We see that the $a$ dependence of $A_{\mu,\rm eff}^a$, derived using
symmetries and given in eq.~\pref{Adirect2}, is supplemented by
an additional discretization error resulting from the
application of the normalization condition at non-zero $a$.

Note that the quantity $Z_A^0$ appears only in the combination  
$Z_{\rm A, Loc}/Z_{\rm A}^0$ in eqs.~\pref{eq:AlocWChPT}
and \pref{ZAWChPT}, and the individual value of $Z_A^0$ is not
necessary. This factor is needed only when expressing the bare
lattice current in the intermediate Symanzik theory. 
In fact, the combination $Z_{\rm A, Loc }/Z_{\rm A}^0$, and also 
the analogue for the vector current, $Z_{\rm V, Loc }/Z_{\rm V}^0$, 
may be interpreted as renormalization constants $Z_{\rm A, eff }$
and $Z_{\rm V, eff }$ in the chiral effective theory \cite{Aoki:2007es}. 

We also remark that the {\em form} of $A_{\mu,\rm ren,Loc,eff}^a$ applies
to any lattice current---local or improved.
These cases only differ in the values of the LECs.
Thus in the following we will drop the subscript ``Loc'' on
the renormalized current in the chiral effective theory.

We close this section by investigating the dependence of
$z_A(t)$ on $t$, $\vec p$ and $\vec q$. There turn out to
be three distinct cases (recalling that $\vec p,\vec q\ne 0$):
\begin{itemize}
\item
$\vec p=\vec q$. This is the simplest case to implement
practically, and leads to $z_A(t)=0$. Thus it turns out that,
in this case,
there are no additional $O(a)$ terms introduced by the current
normalization.
\item
$\vec p$ parallel to $\vec q$. Then, for $|t|\gg 1/|\vec p-\vec q|$,
the product of cosh and exponential becomes $1/2$, and so
$z_A\to 1/2$.
\item
All  other non-vanishing $\vec p$ and $\vec q$.
Here, for $|t|\gg 1/|\vec p-\vec q|$, the exponential
overwhelms the cosh and $z_A\to 1$.
\end{itemize}
We stress, however, that in both the second and third cases $z_A$
depends on $t$ for non-asymptotic values of $t$.

\section{Application: Pion decay constant}
\label{sect:decayconstant}

At this point we have completed the construction and matching of the
effective currents.  Now we can proceed and compute correlation
functions involving these currents. As a simple but important example
we calculate the pion decay constant.  Given the presence of
the $z_A(t)$ contribution, our result
differs, in general, from previously published ones.

\subsection{Decay constant at tree level}

Expanding the renormalized axial-vector current in powers of the pion
fields we obtain to leading order the expression 
\bea 
A_{\mu,\rm ren,eff}^a =
& = & 
i f \frac{\partial_{\mu}\pi^{a}}{\sqrt{Z_\pi}} 
\left( 1 + \frac{4}{f^{2}} \hat{a} 
\left(W_{45}\csw + W_A\ca\right)
\left[2 - z_A(t)\right]\right).
\label{Aretree} 
\eea 
This depends, as required, only on the ``physical'' combination $W_{45}\csw + W_A \ca$
of LECs. We can immediately
read off the following expression for the leading order decay constant: 
\bea
\label{eq:fpiLO}
f_{\pi,{\rm LO}} 
& = & 
f\left(1 + \frac{4}{f^{2}} \hat{a} 
\left(W_{45}\csw + W_A\ca\right)
\left[2 - z_A(t)\right]\right).
\eea 
We see that the O($a$) corrections depend not only on the
form of the underlying lattice action and currents (through the
LECs) but also on the choice of renormalization condition
[through $z_A(t)$]. The $z_A(t)$ term always reduces the size
of these corrections, although by no more than a factor of two
(which happens in
the third case discussed at the end of the previous section).
The decay constant is free of O($a$)
corrections only if both the action and the axial-vector
current are improved, i.e.\ for $\csw=\ca=0$, in accordance 
with what we know from the Symanzik effective theory.

We want to comment on the origin of discrepancies between 
\pref{eq:fpiLO} and previously published results for the decay constant. 
Reference \cite{Rupak:2002sm} finds\footnote{%
  Ref.~\cite{Rupak:2002sm} quotes
  explicitly the 3-flavor result, which we changed to the
  corresponding 2-flavor result to make the comparison.}
\bea\label{Shoreshfpi} 
f_{\pi,{\rm LO}}^{\rm RS} 
& = & 
f\left(1 + \frac{8}{f^2}\hat{a} W_{45}\csw\right)\,, 
\eea 
There is no correction
proportional to $W_A\ca$ since their calculation used the Noether
current as the axial-vector current~\cite{ShoreshPC}.
This misses the O($a$) correction in \pref{SymAC}, and
cannot be correct since the result \pref{Shoreshfpi} is O($a$) improved 
if only the action (and not the current) is improved.

Reference \cite{Sharpe:2004ny} finds the same form
as \pref{eq:fpiLO} except without the $z_A(t)$ contribution.
These authors assumed that a non-perturbative renormalization
condition had been applied, but did not include the impact
of applying the condition at non-vanishing lattice spacing.

\subsection{Decay constant to one loop}

It is straightforward to compute the leading correction to the
tree-level result \pref{eq:fpiLO}.  Expanding the axial-vector current
in eq.\ \pref{Adirect} to higher powers of the pion fields one obtains
the one-loop contributions to $f_{\pi}$. The integrals that appear can be
regularized using dimensional regularization. The counterterms for the
divergences are provided by the tree-level contribution of the NLO
terms in the axial-vector current, c.f.\ \pref{dANLO} and
\pref{dANLO2}. Even though we have not explicitly given all possible
NLO corrections of O($am,ap^{2}$), it is easy to convince oneself that
all contributing tree-level terms with one partial derivative are of
the form $if\partial_{\mu}\pi^{a}\cdot a m$. Expressing $m$ by the
tree-level pion mass according to \pref{eq:Mpitree}, we obtain the
counterterm 
\bea 
A_{\mu}^{a}[a M^{2}_{\pi} ]_{{\rm CT}} 
& = &
if\partial_{\mu}\pi^{a}\tilde{W}_{\rm A3} \hat{a}M^{2}_{\pi}/f^4\,.
\eea 
For simplicity we have absorbed the coefficients $\csw$ and $\ca$,
and the contribution proportional to $z_A(t)$,
in the LEC $\tilde{W}_{\rm A3}$, since in practice these and the
$W$ coefficients are difficult to disentangle.  The additional factor
$1/f^4$ is introduced for convenience, since it leads to a
dimensionless coefficient $\tilde{W}_{\rm A3}$.
 
The final one-loop result for $f_{\pi}$ is then given by
\bea\label{eq:fpiNLO} 
f_{\pi,{\rm 1-loop}} & = & f\left(1 +
\frac{\hat{a}}{f^2} \tilde W_{\rm A1} - \frac{1}{16\pi^{2}f^{2}}\left[1 +
  \frac{\hat{a}}{f^2}\tilde{W}_{\rm A2}\right]
M^{2}_{\pi}\ln\frac{M^{2}_{\pi}}{\mu^{2}} + \frac{8}{f^{2}}
M^{2}_{\pi} \left[L_{45} + \frac{\hat{a}}{f^2}\tilde{W}_{\rm
    A3}\right]\right).\nn\\ && 
\eea 
The coefficients are
$\tilde{W}_{\rm A1} =4(W_{45}\csw + W_A\ca)[2-z_A(t)]$ (as above)
and 
$\tilde{W}_{\rm A2} = 4(W_{45}\csw +W_{A}\ca)[1-z_A(t)]$.
Note that both coefficients depend on the physical combination of LECs, 
as expected. 

A couple of comments concerning \pref{eq:fpiNLO} are in order. In the
continuum limit we reproduce the familiar result for $f_{\pi}$ from
continuum ChPT for $N_{f}=2$ \cite{Gasser:1983yg}. Away from the
continuum limit the result is modified not only
by terms of O($a$), but also by contributions of
O($aM_{\pi}^{2}$) and O($aM_{\pi}^{2}\ln M_{\pi}^{2}$).  

The coefficients $\tilde{W}_{\rm A1}$ and $\tilde{W}_{\rm A2}$ are in general not independent. For example, the first case discussed at the end of the last section has $\tilde{W}_{\rm A1} = 2 \tilde{W}_{\rm A2}$ (for asymptotically large $t$ values). For the third case we even find  $\tilde{W}_{\rm A2} = 0$, so the coefficient of the chiral logarithm is free of O($a$) artifacts. Since in this case also $\tilde{W}_{\rm A1}$ assumes its minimal value this is the theoretically preferred renormalization condition.

Except for the special case with $\tilde{W}_{\rm A2} = 0$
the coefficient of the chiral logarithm receives an O($a$)
correction in the form of the factor 
$[1 + \hat{a}\tilde{W}_{\rm A2}/f^2]$. 
Consequently, the coefficient of the chiral logarithm
is, in contrast to continuum ChPT, not a universal coefficient
depending on $f$ (and $N_{f}$) only, but on the (non-universal)
lattice artifacts too. This fact has previously been stressed in
Ref.~\cite{Aoki:2003yv}.

Note that the combination $L_{45}$ appears effectively in the lattice spacing
dependent combination $L_{45}^{\rm eff}(a)=L_{45} + \hat{a}
\tilde{W}_{\rm A3}/f^2$. Determinations of $L_{45}$ based on
simulations at one lattice spacing only are potentially dangerous
because the size of the contribution $\hat{a}\tilde{W}_{\rm A3}/f^2$
is a priori unknown.

\section{Concluding remarks}
\label{sect:conclusion}

In this paper we have reconsidered the construction of the vector and axial
vector currents in WChPT. Due to the explicit chiral symmetry breaking
in Lattice QCD with Wilson fermions two aspects need to be taken into
account which are not present in continuum ChPT.

Firstly, the local lattice currents are not conserved, and in general
they do not map onto the conserved currents in WChPT. In particular,
the WChPT currents are not the Noether currents associated with the
chiral symmetries, not even at leading order in the chiral
expansion. The reason is that the currents in the Symanzik theory have
O($a$) corrections which are not related to the effective action.

Secondly, the matching of the currents needs to take into account the
finite renormalization of the local lattice currents. In order to
properly match the currents the same renormalization conditions that
have been employed for the lattice currents must be imposed on the
WChPT currents. 

Depending on the particular choice for the renormalization conditions
the expressions for the renormalized currents differ by terms of
O($a$). As a result, the WChPT predictions for matrix elements of the
currents are different as well. Consequently, a result for an
observable like $f_{\pi}$ should make reference to the renormalization
condition one has adopted. This has to be so at some level, since the
lattice data differs too depending on the condition one has chosen.
What we find is that the dependence enters at O($a$).

At a technical level, our result does not change the number of
low-energy coefficients that enter into WChPT predictions at O($a$).
In particular, as stressed in Ref.~\cite{Sharpe:2004ny}, there are
only two combinations of the LECs that can enter into physical
quantities, allowing their O($a$) corrections to be related.
What changes is the nature of these relations, which now depend
to the choice of normalization condition.

The considerations of this paper apply more generally.
Another example from WChPT is the ratio of the 
renormalization factors for pseudoscalar and scalar densities.
This must be determined non-perturbatively, e.g. by enforcing 
Ward identities, and will presumably receive O($a$) corrections
similar to those for the axial current, although we have not worked
out the details. The combination $m S$ is, however, protected by
exact lattice WTIs.

More generally still, our results emphasize the (perhaps rather
obvious) point that non-perturbative renormalization conditions
generically introduce additional discretization errors.
These must be (and in practice usually are being)
accounted for when extrapolating to the continuum limit.

\section*{Acknowledgments}

O. B.\ acknowledges useful discussions with Johan Bijnens, Maarten
Golterman and Rainer Sommer. We also thank Rainer Sommer for
feedback on a first draft of this paper.

This work is supported in part by the Grants-in-Aid for Scientific
Research from the Ministry of Education, Culture, Sports, Science and
Technology (Nos. 20340047,20105001,20105003), by the Deutsche
Forschungsgemeinschaft (SFB/TR 09) and by the U.S. Department of
Energy.

\begin{appendix}

\section{The generating functional method}
\label{sourcemethod}

In section \ref{ssect:MatchingtoChPT} we constructed the effective
currents in a rather direct way. We first wrote down the most general
currents that transform as vector and axial-vector currents. In a
second step we imposed appropriate Ward identities that these currents
must obey, and this led to various constraints on the LECs in these
currents.

We already mentioned at the end of section \ref{ssect:MatchingtoChPT}
that this is not the way one proceeds in continuum ChPT
\cite{Gasser:1983yg,Gasser:1984gg}. Instead, there one sets up the
generating functional for correlation functions involving the currents
(and the scalar and pseudoscalar density), and matches this to the
analogous generating functional in the chiral effective theory. In
this way the currents are obtained by functional derivatives with
respect to the sources.

A crucial link in the matching is the invariance under local chiral
transformations, which plays the role of a gauge symmetry. The local
invariance implies that the sources enter the chiral Lagrangian in a
very restricted way, namely in form of a gauge-covariant derivative.

One may ask if the same procedure is also possible once we include the
lattice spacing corrections. In order to answer this question we
derive in detail the generating functional for the Symanzik effective
theory and match it to the one in WChPT. We will see that we indeed
obtain the same currents as the ones in \pref{VeffNoether} and
\pref{Adirect2}. However, we will also see that it is more complicated
to maintain the invariance under local chiral transformations.

The ``generating functional method'' has been used before in WChPT
in Ref.\ \cite{Sharpe:2004ny}. However, as discussed
in Ref.\ \cite{SharpeNara}, the generating functional 
of Ref.\ \cite{Sharpe:2004ny} violates the
invariance under local chiral transformations at nonzero lattice
spacing; only the leading (continuum) part respects this
symmetry. Consequently, the construction of the axial-vector current
requires some care: part of the O($a$) corrections have to be mapped
separately into the effective theory \cite{SharpeNara}. The final
result found in \cite{SharpeNara}
for the currents turns out, however, to have the form 
given in Ref.\ \cite{Sharpe:2004ny},
as we derive here in a systematic way.

\subsection{Generating functional in continuum QCD and ChPT} 
\label{sect_Cont_GF}

In order to prepare our discussion we first give a brief review of the 
construction in continuum ChPT, mainly to introduce our notation.  We
start by defining a (Euclidean) QCD Lagrangian including a part that
includes sources for the currents and densities, 
\bea
\label{FullLag}
\Lagno & = & \Lag{QCD} -i \Lag{Source}\,, 
\eea 
where $\Lag{QCD}$ is
the usual massless QCD Lagrangian, while the second part
\bea
\label{sourcetermcont} 
\Lag{Source} & = & \psibar \gamma_{\mu}
\left[v_{\mu}(x) + \gamma_{5} a_{\mu}(x)\right] \psi + \psibar\left[
  s(x) + \gamma_{5} p(x)\right]\psi 
\eea 
contains sources for the
vector and axial-vector currents ($v_{\mu}, a_{\mu}$) and for the
scalar and pseudoscalar densities ($s, p$).  These sources are matrix
valued fields, given by \bea v_{\mu}(x) & = & v^{a}_{\mu}(x)
T^{a},\qquad a_{\mu}(x)\, =\, a^{a}_{\mu}(x)T^{a},\nn\\ s(x) &=&
s^a(x)T^a\,,\qquad p(x)\,=\,p^a(x)T^a\,, \eea where $T^{a}$ are the
hermitian SU($N_{f}$) generators, normalized according to
$\mbox{tr}(T^{a}T^{b}) = \delta^{ab}/2$. We are interested in
is $N_f=2$, for which $T^{a} = \sigma^{a}/2$, with
$\sigma^{a}$ the usual Pauli matrices.

After integrating over space time we obtain the action in the presence
of the sources, $S= \int d^4x \Lagno(x)$. By taking functional
derivatives with respect to the sources we obtain the vector and axial
vector current and the scalar and pseudoscalar densities: 
\bea
i\frac{\delta S}{\delta v_{\mu}^{a}(x)} & = & \psibar(x) \gamma_{\mu}
T^{a} \psi(x)\,=\, V_{\mu}^{a}(x)\,,\label{Vcontder}\\ 
i\frac{\delta
  S}{\delta a^{a}_{\mu}(x)} & = & \psibar(x) \gamma_{\mu} \gamma_{5}
T^{a} \psi(x)\,=\, A_{\mu}^{a}(x)\,,\label{Acontder}\\ 
i\frac{\delta
  S}{\delta s^{a}(x)} & = & \psibar(x) T^{a} \psi(x)\,=\,
S^{a}(x)\,,\\ 
i\frac{\delta S}{\delta p^{a}(x)} & = & \psibar(x)
\gamma_{5} T^{a} \psi(x)\,=\, P^{a}(x)\,.  
\eea 
The key observation is
that the Lagrangian \pref{FullLag} is invariant under {\em local}
$SU(N_{f})_{R} \times SU(N_{f})_{L}$ transformations, which act on the
fermion fields according to 
\bea 
\psi(x) &\rightarrow &
\psi^{\prime}(x) \,=\, R(x) P_{+} \psi(x) + L(x) P_{-}
\psi(x),\nn\\ \psibar(x) &\rightarrow & \psibar^{\prime}(x) \,=\,
\psibar(x) P_{+} L^{\dagger}(x) + \psibar(x) P_{-} \Rdag(x)\,.  
\eea
The projectors are defined in the usual way,
$P_{\pm}=(1\pm\gamma_5)/2$, 
and project onto fields with definite chirality, 
\bea 
\psi_{R} & = & P_{+}\psi,\qquad \psi_{L} \,=\, P_{-}
\psi\,,\nn \\ \psibar_{R} & = & \psibar P_{-} ,\qquad \psibar_{L}
\,=\, \psibar P_{+}\,.  
\eea 
Crucial for the local invariance is the
non-trivial transformation of the source fields 
\bea 
v_{\mu}+ a_{\mu}
& \rightarrow & v_{\mu}^{\prime} + a_{\mu}^{\prime} \,=\, R(v_{\mu} +
a_{\mu})\Rdag + iR\partial_{\mu}\Rdag \label{trafocurrentright}
\,,\\ 
v_{\mu}- a_{\mu} & \rightarrow & v_{\mu}^{\prime} -
a_{\mu}^{\prime} \,=\, L(v_{\mu} - a_{\mu})\Ldag +
iL\partial_{\mu}\Ldag \label{trafocurrentleft} \,,\\ 
s + p &
\rightarrow & s^{\prime} + p^{\prime} \,=\, L(s+p)\Rdag\,,\\ 
s - p &
\rightarrow & s^{\prime} - p^{\prime} \,=\, R(s-p)\Ldag\,.  
\eea 
(For notational simplicity we drop from now on the argument $x$).  
The invariance is more easily seen if we express the Lagrangian in terms
of left- and right-handed fields, 
\bea
\label{SourceLR} 
\Lag{Source} &= & 
\psibarR \gamma_{\mu} \left[v_{\mu} + a_{\mu}\right] \psiR +
\psibarL \gamma_{\mu} \left[v_{\mu} - a_{\mu}\right] \psiL + \psibarL
\left[s +p\right] \psiR + \psibarR \left[s -p\right] \psiL\,, 
\eea
which obviously is invariant under global transformations. Note that
global transformations leave $\Lag{QCD} $ and $\Lag{Source}$
independently invariant. Under local transformations the derivative
part of $\Lag{QCD} $ produces extra terms which are cancelled by the
derivative terms in \pref{trafocurrentright} and
\pref{trafocurrentleft}. The reason why this cancellation works is
related to the fact that the currents are the conserved Noether
currents associated with chiral transformations and hence stem from
local variations of the derivative term in the Lagrangian.

The Lagrangian $\Lagno$ can be conveniently rewritten in a form that
makes the local invariance more transparent. Firstly, the source term
in \pref{SourceLR} suggests to introduce sources for right- and
left-handed currents, 
\bea\label{defrightleftsources} 
r_{\mu} & = &
v_{\mu}+ a_{\mu}\,,\qquad l_{\mu} \, = \, v_{\mu}- a_{\mu}\,, \eea
which transform according to \bea r_{\mu} & \rightarrow &
r_{\mu}^{\prime} \,=\, Rr_{\mu}\Rdag +
iR\partial_{\mu}\Rdag \label{trafocurrentright2} \,,\nn\\ 
l_{\mu} 
&\rightarrow & l_{\mu}^{\prime} \,=\, Ll_{\mu}\Ldag +
iL\partial_{\mu}\Ldag \label{trafocurrentleft2} \,.  
\eea 
Vector and axial-vector currents are then obtained by 
\bea 
V_{\mu}^{a}(x)& = & i
\left[ \frac{\delta S}{\delta r_{\mu}^{a}(x)} + \frac{\delta S}{\delta
    l_{\mu}^{a}(x)} \right] \,,\label{LRDerivV}\\
A_{\mu}^{a}(x)& = &
i\left[ \frac{\delta S}{\delta r_{\mu}^{a}(x)} - \frac{\delta
    S}{\delta l_{\mu}^{a}(x)} \right] \label{LRDerivA}\,.  
\eea 
The right- and left-handed sources allow the definition of covariant
derivatives 
\bea 
D_{\mu}^{R} & = & D^{\rm color}_{\mu} - i
r_{\mu}\,,\qquad D_{\mu}^{L} \, = \, D^{\rm color}_{\mu} - i
l_{\mu}\,,\label{contcovder} 
\eea 
where $D^{\rm color}_{\mu}$ is the
covariant derivative with respect to local $SU(3)$ color
transformations. By construction, $D_{\mu}^{R}\psiR, D_{\mu}^{L}\psiL$
transform as the chiral fields themselves, and $\Lagno$ is simply
given by 
\bea
\label{LagcovDer} 
\Lagno & =& \psibarR \gamma_{\mu}
D_{\mu}^{R} \psiR + \psibarL \gamma_{\mu} D_{\mu}^{L} \psiL +
\Lag{Gauge}\,, 
\eea 
where $\Lag{Gauge}$ is the gauge field part of the
QCD Lagrangian containing the gluon fields $G_{\mu}^a(x)$. Obviously,
\pref{LagcovDer} is invariant under local chiral transformations.

Having $\Lagno$ in hand we can now define a generating functional for
correlation functions involving the currents and densities, 
\bea
Z_{\rm QCD}[r_{\mu},l_{\mu},s,p] & =& \frac{1}{Z_0} \int
{\mathcal{D}}[G_{\mu},\bar{\psi},\psi] e^{-\int
  d^4x\,\Lagno[r_{\mu},l_{\mu},s,p]}\,,\label{defZQCD} 
\eea
 where $Z_0$ denotes the partition function, i.e.\ $ Z_{\rm QCD}$ for
vanishing sources.\footnote{%
  Strictly speaking, the scalar source
  should not vanish but be set to the physical quark mass.}  
Taking functional derivatives with respect to the sources one generates all
possible correlation functions involving the currents and densities;
the generating functional for connected correlation functions can be
defined as usual by the logarithm of $ Z_{\rm QCD}$.
 
The matching to the chiral effective theory is now done by requiring
that the generating functional in the chiral effective theory is the
same as the one in the fundamental theory, 
\bea 
Z_{\rm   QCD}[r_{\mu},l_{\mu},s,p] &=& Z_{\rm
  chiral}[r_{\mu},l_{\mu},s,p]\,,\label{MatchingZ}\\ 
Z_{\rm   chiral}[r_{\mu},l_{\mu},s,p] & = & \frac{1}{Z_{\rm 0} } \int
{\mathcal D}[\pi] e^{- \int d^{4}x \, {\cal L}_{\rm
    chiral}[r_{\mu},l_{\mu},s,p]}.\label{Zchiraldef} 
\eea 
The equality in \pref{MatchingZ} is meant 
in the sense that the two sides coincide
order by order in a low energy expansion, where the long-distance
correlation functions are dominated by the pion pole
\cite{Leutwyler:1993iq}. Correlation functions in the effective theory
are obtained by the same functional derivatives as in the underlying
theory. Provided the generating functionals, as functions of the
sources, are the same, arbitrary correlation functions also
agree. Moreover, the QCD symmetries are carried over to the effective
theory, i.e.\ the Ward identities of QCD are correctly reproduced by
the effective theory. Consequently, the right hand side must also be
invariant under local chiral transformations, and this invariance
provides one constraint in the construction of 
${\cal L}_{\rm   chiral}$. 
Other constraints are provided by parity ($P$) and charge
conjugation ($C$).

The effective Lagrangian ${\cal L}_{\rm chiral}[r_{\mu},l_{\mu},s,p]$
is obtained in a systematic low energy expansion, and it has been
derived by Gasser and Leutwyler through next-to-leading order
\cite{Gasser:1983yg,Gasser:1984gg}. We do not repeat their result here
but emphasize that the local chiral invariance implies that the
sources for the right and left handed currents can only enter the
effective Lagrangian through the covariant derivative
\bea
\label{CovDer} 
D_{\mu}\Sigma & =& \partial_{\mu}\Sigma + i \Sigma
r_{\mu} - i l_{\mu} \Sigma 
\eea 
on the usual chiral field $\Sigma$,
and through the field strength tensors 
\bea
\label{ContFST} 
r_{\mu\nu}
& = & \partial_{\mu}r_{\nu} - \partial_{\nu}r_{\mu} + i
[r_{\mu},r_{\nu}],\qquad l_{\mu\nu} \,= \, \partial_{\mu}l_{\nu} -
\partial_{\nu}l_{\mu} + i [l_{\mu},l_{\nu}]\,.  
\eea 
The
transformation behavior of these objects is as expected, taking into
account $\Sigma \rightarrow L\Sigma R^{\dagger}$, 
\bea 
D_{\mu}\Sigma &
\rightarrow & L (D_{\mu}\Sigma) R^{\dagger}\,,\qquad r_{\mu\nu} \,
\rightarrow \,R r_{\mu\nu} R^{\dagger}\,,\qquad l_{\mu\nu } \,
\rightarrow \, L l_{\mu\nu} L^{\dagger}\,.  
\eea

\subsection{Generating functional in the Symanzik effective theory}
\label{sect_Cont}
Following the development in the continuum, we want to define a generating
functional in the Symanzik effective theory, which may then be matched
to the chiral effective theory. As before we want to obtain the
currents and densities by taking derivatives with respect to the
source fields. As we will see this is not as straightforward as in
continuum QCD. For simplicity we will deal here only with the currents
and ignore the densities. This reveals the main obstacles in the
procedure.

As in the continuum case we want to define a source term in the
Lagrangian, such that taking the derivatives with respect to sources
produce the Symanzik currents, given in \pref{SymVC} and
\pref{SymAC}. In addition, we want to do this in a way that maintains
invariance under local chiral transformations, since this symmetry
provides one of the links in the matching to the chiral effective
theory.

It is illustrative to first show that the naive generalization of the
continuum case fails in this respect. Suppose we write down a source
term that directly couples source fields to the vector and axial
vector current, 
\bea 
\Lag{Source,{\rm Sym}} &=& v^a(x)V^a_{\mu,{\rm
    Sym,Loc}}(x) + a^a_{\mu}(x)A^a_{\mu,{\rm Sym, Loc}}(x)\,.  
\eea
Just as in the continuum---cf.\ eq.\ \pref{Vcontder} and
\pref{Acontder}---functional derivatives with respect to $v^a(x),
a^a(x)$ produce the desired currents. However, local invariance is
lost: The O($a$) corrections in the currents and the derivative terms
in \pref{trafocurrentright}, \pref{trafocurrentleft}, generate O($a$)
terms under local chiral transformations that are not cancelled by the
variation of the Symanzik effective action in
\pref{eq:SymAction}. That is of course no surprise, it simply reflects
the fact that chiral symmetry is explicitly broken in the Symanzik
effective theory and that the currents in \pref{SymVC} and
\pref{SymAC} are not Noether currents associated with exact chiral
symmetries.

In order to maintain local chiral invariance we proceed differently
and introduce separate sources fields for the O($a$) corrections in
the currents. Our source Lagrangian has the form 
\bea 
\Lag{Source,{\rm Sym}} &=& 
\Lag{Source, {\rm ct}} + \Lag{Source, a}\,.  
\eea 
The first term with subscript ``ct'' is the familiar one from continuum
QCD, defined in \pref{sourcetermcont}. As before, one can rewrite it
in terms of right and left handed sources using
\pref{defrightleftsources}. 
The second part reads 
\bea 
\Lag{Source, a}
& = & \phantom{+} (D^{L}_{\nu}\psibarL) i\sigma_{\mu\nu}
C_{V}r_{V,\mu} \psiR+ \psibarL i\sigma_{\mu\nu} l_{V,\mu} C_{V}
D^{R}_{\nu}\psiR\nonumber\\ 
& & + (D^{R}_{\nu}\psibarR)
i\sigma_{\mu\nu} C_{V}^{\dagger}l_{V,\mu}\psiL + \psibarR
i\sigma_{\mu\nu} r_{V,\mu}C_{V}^{\dagger}D^{L}_{\nu}\psiL \nonumber
\\ 
& & + (D^{L}_{\mu}\psibarL)\gamma_{5} C_{A} r_{A,\mu} \psiR +
\psibarL \gamma_{5}l_{A,\mu} C_{A} D^{R}_{\mu}\psiR\nonumber\\ 
&& +
(D^{R}_{\mu}\psibarR) \gamma_{5} C_{A}^{\dagger} l_{A,\mu} \psiL +
\psibarR \gamma_{5} r_{A,\mu}C_{A}^{\dagger}
D^{L}_{\mu}\psiL\,.\label{Lsourcea} 
\eea 
This is essentially nothing
but a source term for the O($a$) corrections to the currents as given
in \pref{Vmuacorr} and \pref{Amuacorr}. We have introduced sources
$r^a_{X, \mu}, l^a_{X, \mu},\, X = V$ or $A$, which transform
according to 
\bea 
r_{X, \mu} \, \rightarrow \, r_{X,\mu}^{\prime}
\,=\, Rr_{X,\mu}\Rdag &\quad &l_{X,\mu} \,\rightarrow\,
l_{X,\mu}^{\prime} \,=\, Ll_{X,\mu}\Ldag \label{trafosourceXleft} \,.
\eea 
Note that this is the transformation law under {\em local}
transformations even though there are no terms involving derivatives
of $L$ and $R$.  Note also that $\Lag{Source, a}$ contains the
covariant derivatives $D_{\mu}^{R},D_{\mu}^{L}$, defined in
\pref{contcovder}, which involve the sources $l_{\mu},r_{\mu}$ for the
continuum parts in the Symanzik currents.

It is easily checked that the source term \pref{Lsourcea} is invariant
under local chiral transformations (taking into account the
transformation laws for $C_V, C_A$ given in
\pref{TrafoCs}). Therefore, also the total Lagrangian
\bea
\label{FullLagSym} 
\Lagno & = & \Lag{Sym} -i \Lag{Source, Sym}
\eea 
is locally invariant. The currents are then obtained from 
$S=\int d^4x \Lagno(x)$ according to 
\bea 
V_{\mu,{\rm Sym}}^{a}(x)& = &
i\left[ \frac{\delta S}{\delta r_{\mu}^{a}(x)} + \frac{\delta
    S}{\delta l_{\mu}^{a}(x)} + \frac{\delta S}{\delta
    r_{V,\mu}^{a}(x)} + \frac{\delta S}{\delta l_{V,\mu}^{a}(x)}
  \right] \,,\label{LRDerivVSym}\\ 
A_{\mu,{\rm Sym}}^{a}(x)& = &
i\left[ \frac{\delta S}{\delta r_{\mu}^{a}(x)} - \frac{\delta
    S}{\delta l_{\mu}^{a}(x)} + \frac{\delta S}{\delta
    r_{A,\mu}^{a}(x)} - \frac{\delta S}{\delta l_{A,\mu}^{a}(x)}
  \right]\label{LRDerivASym}\,, 
\eea 
which is a generalization of the
prescription in the continuum, \pref{LRDerivV} and
\pref{LRDerivA}. Note that we implicitly assume here that the spurion
fields $A, C_V,C_A$ are set to their ``physical'' values $a\csw,
a\cv,a\ca$ after the derivatives have been taken.

So far we focused on the invariance under local chiral
transformations. For the matching to the chiral effective theory the
discrete symmetries $P$ and $C$ are also needed. In addition, there is
one property in the source term that we also have to preserve. The
source term in \pref{Lsourcea} depends on the sources and the spurion
fields $C_V$ and $C_A$. The dependence is such that the vector current
sources $r_V,l_V$ only couple to $C_V$ and the axial-vector sources
only to $C_A$. A mixed term, involving $C_Ar_{V,\mu}$ for example, is
not present. It is mandatory to preserve this feature, otherwise the
vector current in the effective theory may end up with a contribution
proportional to $\ca$, which is certainly not the case.  In order to
achieve this we choose to generalize the source term without changing
the results for the currents derived from it.

Instead of one spurion field $C_V$ we introduce two of them, $C_{V,1}$
and $C_{V,2}$, and make the replacements 
\bea 
C_{V}r_{V,\mu} &
\longrightarrow & C_{V,1}r_{V,\mu}\,\qquad l_{V,\mu}
C_{V}\,\longrightarrow\,l_{V,\mu} C_{V,2}\,.  
\eea 
The symmetry
properties of $\Lag{Source,a}$ remains unchanged if both $C_{V,1}$ and
$C_{V,2}$ transform as $C_V$, and if both have the same ``physical''
value $a\cv$. However, the source term is now invariant under more
general symmetry transformations, namely 
\bea 
r_{V,\mu} \,
\longrightarrow \, H_V r_{V,\mu} R^{\dagger},& \quad & C_{V,1}
\,\longrightarrow \, LC_{V,1}H_V^{\dagger}\,,\nn\\ 
l_{V,\mu} \,
\longrightarrow \, L l_{V,\mu} G_V^{\dagger},& \quad & C_{V,2}
\,\longrightarrow \, G_VC_{V,2}R^{\dagger}\,,\label{hiddenSym1} 
\eea
where $H_V$ and $G_V$ are two independent local $SU(2)$ matrices. The
origin of these two {\em hidden local symmetries} is the fact that the
right and left-handed sources and the spurions $C_{V,i}$ only appear
next to each other in the source term. Note that we recover the
previously discussed transformation laws with $H_V=R$ and $G_V=L$.

The same generalization can be done with the axial-vector part in the
source term after we have introduced $C_{A,1}$ and $C_{A,2}$ and
postulated the transformation behavior 
\bea 
r_{A,\mu} \,
\longrightarrow \, H_A r_{A,\mu} R^{\dagger},&\qquad &C_{A,1}
\,\longrightarrow \, LC_{A,1}H_A^{\dagger}\,,\nn\\ l_{A,\mu} \,
\longrightarrow \, L l_{A,\mu} G_A^{\dagger},&\qquad & C_{A,2}
\,\longrightarrow \, G_AC_{A,2}R^{\dagger}\,.\label{hiddenSym2} 
\eea
Since the symmetries $H_A,G_A$ are independent of the ones in the
vector current part we can no longer form invariants with vector
current sources and axial-vector spurions $C_{A,i}$ and 
{\em vice-versa}. 
This will become important in the next section where we
construct the currents in the chiral effective theory.

In analogy to \pref{defZQCD} we can now define a generating functional
in the Symanzik effective theory. The number of external fields,
however, has grown significantly, 
\bea
\label{SymGenFunct} 
Z_{\rm Sym}&=&Z_{\rm
  Sym}[r_{\mu},l_{\mu},r_{V,\mu},l_{V,\mu},r_{A,\mu},l_{A,\mu},
  A,C_{V,1},C_{V,2},C_{A,1}, C_{A,2}] \,, 
\eea 
and here we still have
not included any sources associated with the scalar and pseudoscalar
density. Apparently, the method of constructing correlation functions
via a generating functional looses its simplicity away from the
continuum limit. In the naive continuum limit \pref{SymGenFunct}
reduces to $Z_{\rm QCD}$ with its dependence on $r_{\mu},l_{\mu}$
only.

\subsection{Matching to WChPT}

Similarly to the continuum case we now match the Symanzik effective
theory to Wilson ChPT by demanding that the generating functionals in
both theories agree 
\bea
\label{MatchingSymWChPT} 
Z_{\rm Sym}&=& Z_{\rm chiral}\,, 
\eea 
where the right hand side is defined in analogy to
\pref{Zchiraldef}. For simplicity we have dropped the dependence on
the large number of external fields, specified in \pref{SymGenFunct}.

The effective Lagrangian entering the right hand side in
\pref{MatchingSymWChPT} is build in terms of the pseudoscalar field
$\Sigma$ and its covariant derivative $D_{\mu}\Sigma$ as well as all
external fields. Many terms have already been constructed and can be
found in the literature. Firstly, one finds the familiar
Gasser-Leutwyler Lagrangian \cite{Gasser:1983yg,Gasser:1984gg}, which
provides the continuum part. Terms involving the spurion field $A$ 
are given in Ref.~\cite{Sharpe:2004ny}.
Hence we focus here
 on the new contributions stemming from the O($a$) terms in the
currents, i.e.\ those involving the fields
$r_{V,\mu},l_{V,\mu},r_{A,\mu},l_{A,\mu},C_{V,i},C_{A,i}$.

It will be useful to introduce the following combinations: 
\bea
S_{V,\mu} & = & i(C_{V,1}r_{V,\mu} + l_{V,\mu}C_{V,2}),\qquad
S_{A,\mu} \, = \,i(C_{A,1}r_{A,\mu} + l_{A,\mu}C_{A,2})\,,\nn
\\
A_{V,\mu} & = & i(C_{V,1}r_{V,\mu} - l_{V,\mu}C_{V,2}),\qquad
A_{A,\mu} \, =\, i(C_{A,1}r_{A,\mu} -
l_{A,\mu}C_{A,2}).\label{Symmetrizedsources} 
\eea 
These combinations
are automatically invariant under the four hidden symmetries involving
$H_X,G_X$ with $X=V,A$. The factor $i$ is just convention, inspired by
the observation that the terms involving the continuum sources also
always include an $i$. For convenience we have summarized the
transformation behavior of these quantities in table
\ref{table:Spurions}. We also list the transformation rules under
parity and charge conjugation, which one also needs for the
construction of invariants under all symmetries.

\begin{table}[t]
\begin{tabular}{|c|c|c|c|} \hline
Field & Chiral & Charge Conj. $C$ & Parity $P$ \\ \hline 
$\Sigma$ & $L \Sigma L^{\dagger}$ & $\Sigma^{T}$ & $\Sigma^{\dagger}$
\\ 
$D_{\mu}\Sigma$ & $LD_{\mu} \Sigma R^{\dagger}$ & $(D_{\mu} \Sigma)^{T}$ &
$(-1)^{\mu}(D_{\mu} \Sigma)^{\dagger}$ 
\\ 
$A$ & $LAR^{\dagger}$ &
$A^{T}$ & $A^{\dagger}$ 
\\ 
$r_{X,\mu}$ & $H_X r_{X,\mu} R^{\dagger}$ &
$-l_{X,\mu}^{ T}$ & $(-1)^{\mu} l_{X,\mu}$ 
\\ 
$l_{X,\mu}$ & $L
r_{X,\mu}G_X^{\dagger}$ & $-r_{X,\mu}^{ T}$ & $(-1)^{\mu} r_{X,\mu}$
\\ 
$C_{X,1}$ & $L C_{X} H_X^{\dagger}$ & $C_{X,1}^{T}$ &
$C_{X,1}^{\dagger}$ 
\\ 
$C_{X,2}$ & $G_X C_{X} R^{\dagger}$ &
$C_{X,2}^{T}$ & $C_{X,2}^{\dagger}$ 
\\ 
$S_{X,\mu}$ & $L S_{X,\mu}
R^{\dagger}$ & $-S_{X,\mu}^{T}$ & $-(-1)^{\mu}S_{X,\mu}^{\dagger}$
\\ 
$A_{X,\mu}$ & $L A_{X,\mu} R^{\dagger}$ & $A_{X,\mu}^{T}$ &
$(-1)^{\mu}A_{X,\mu}^{\dagger}$ 
\\ \hline
\end{tabular}
\caption{Summary of transformation properties. The shorthand notation
  $(-1)^{\mu}$ in the vector quantities represents the sign flip in
  the spatial components under parity.}
\label{table:Spurions}
\end{table}

We are now in the position to construct the terms of O($a$) that will
contribute to the currents. For this we need invariants involving one
power of the sources in \pref{Symmetrizedsources}. Lorentz invariance
then requires one covariant derivative $D_{\mu}$. The simplest
invariants that can be formed are 
\bea 
&& \Tr{D_{\mu}\Sigma
  A_{X,\mu}^{\dagger} + A_{X,\mu}(D_{\mu}\Sigma)^{\dagger} }\,,\qquad
X\,=\,V,A\,.\label{addInvariants} 
\eea 
Analogous invariants involving
$S_{X,\mu}$ cannot be build since they violate charge conjugation.

We do not need to consider any invariants involving
$D_{\mu}S_{X,\mu},D_{\mu}A_{X,\mu}$ or terms quadratic in $S_{X,\mu}$
or $A_{X,\mu}$. Even though there exist non-vanishing invariants they
will not give contributions to the vector or axial-vector current. The
reason is that all these terms contain more than one source field and
derivatives of $C_{X}$. Setting in the end the sources to zero and
$C_{X}$ to its constant final value all these terms vanish.  We also
do not consider the terms involving 
$\Tr{\Sigma^{\dagger}
  D_{\mu}\Sigma \pm \Sigma(D_{\mu}\Sigma)^{\dagger}}$ 
since 
\bea
\mbox{tr }(\Sigma^{\dagger} D_{\mu}\Sigma) & = & \partial_{\mu}
\ln\det \Sigma \,=\, 0\,.  
\eea 
It turns out that the two terms given
in \pref{addInvariants} are the only independent invariants at the
order we are interested in.

In the last section we stressed the importance of the local hidden
symmetries for the correct chiral Lagrangian. If instead of
\pref{hiddenSym1}, \pref{hiddenSym2} we assume the weaker
transformation laws \pref{trafosourceXleft} and \pref{TrafoCs}, we can
construct the additional invariants 
\bea 
&&\Tr{\Sigma^{\dagger}
  D_{\mu}\Sigma r_{X,\mu} + \Sigma (D_{\mu}\Sigma)^{\dagger}
  l_{X,\mu}} \Tr{C_{X}\Sigma^{\dagger} + \Sigma C_{X}^{\dagger}} 
\eea
These invariants require to ``split'' the right and left handed source
fields from the O($a$) spurions $C_{X}$ and build products of two
trace terms, something that is forbidden by the transformation rule
\pref{hiddenSym1}, \pref{hiddenSym2}.

It is straightforward to compute the O($a$) corrections to the
currents following from \pref{addInvariants}. In the case of the
vector current the correction vanishes.  For the axial-vector current
we find the correction 
\bea 
&& 2 a \ca \partial_{\mu} \Tr{T^{a}
  (\Sigma - \Sigma^{\dagger})} 
\eea 
In order to derive the complete
currents including all O($a$) corrections we also need other terms
which have already been given by Ref.~\cite{Sharpe:2004ny}. 
Carrying over their notation, the relevant terms
in the Lagrangian are 
\bea 
\Lagno&=& \frac{{f^2 }}{4}\left\langle
D_{\mu} \Sigma (D_{\mu} \Sigma)^\dagger \right\rangle
+W_{45}\Tr{D_{\mu} \Sigma (D_{\mu} \Sigma)^\dagger}
\Tr{A\Sigma^{\dagger} + \Sigma A^{\dagger}}\nn\\ 
& & + W_{10}
\Tr{D_{\mu} \Sigma (D_{\mu} A)^{\dagger} + D_{\mu} A (D_{\mu}
  \Sigma)^{\dagger}} + 2W_A \Tr{D_{\mu}\Sigma A_{A,\mu}^{\dagger} +
  A_{A,\mu}(D_{\mu}\Sigma)^{\dagger} }
\label{Lagalla}\,.
\eea 
The last term, coming with a new LEC $W_A$, is the one we found
above. We do not include the term with $X=V$ since its contribution to
the vector current vanishes anyway.
   
Before deriving the currents we are free to make a field redefinition
in order to simplify $\Lagno$. Following Ref.~\cite{Sharpe:2004ny} we
perform the change $\Sigma\rightarrow \Sigma + \delta \Sigma$ with
\bea
\label{changeofvar} 
\delta \Sigma &= & \frac{4 W_{10}}{f^{2}}
(\Sigma A^{\dagger} \Sigma - A) 
\eea 
and obtain $\Lagno$ in
\pref{Lagalla} with the modified coefficient $W_{45}\rightarrow
W_{45}+W_{10}$. 
Therefore, without loss we can drop the $W_{10}$ term
in \pref{Lagalla}, as long as we consider only physical quantities.

In fact, if we work only to linear order in the sources,
it is easy to see that the $W_A$ and $W_{10}$ terms in \pref{Lagalla}
are proportional, so that $W_A$ can also be absorbed into
$W_{45}$ by a change of variables~\cite{SharpeNara}.
This holds as long as we consider only physical quantities {\em and}
if all currents are placed at different space-time points.
In order to avoid the latter restriction, and maintain generality,
we do not make this change of variables.
Nevertheless, for the quantities we calculate 
in this paper, which are both physical and involve separated currents, 
the possibility of this change of variables implies
that the LECs must enter in the combination~\cite{Sharpe:2004ny}
$W_{45}\, \csw + W_A \,\ca $.
This provides a check on results.

We can now compute the vector and axial-vector current according to
\pref{LRDerivVSym}, \pref{LRDerivASym}, and find exactly the same
expressions given before in eqs.\ \pref{VeffNoether} and
\pref{Adirect2}.  We therefore conclude that we obtain identical
results for the currents with the generating functional as with our
``direct method''.

\section{Computation of $\mathbf{Z_{\rm A,Loc}}$.}
\label{app_WTI}

In this appendix we derive the result
\pref{ZAWChPT} for the
renormalization constant $Z_{\rm A,Loc}$, which follows from imposing the
Ward-Takahashi identity (WTI) in \pref{AVWI3}.  It will be useful to
first show that \pref{AVWI3} is indeed an identity in (massless)
continuum ChPT. The generalization to WChPT is then straightforward.

We start by establishing 
\beqa
\label{WTIgen} 
\int d\vec x
\epsilon^{abc} \langle \left\{ A_{0,\rm LO}^a(y_0+t,\vec x) 
- A_{0,\rm LO}^a(y-t,\vec x)\right\} 
A_{0,\rm LO}^b(y) {\cal O}_{\rm out} \rangle 
&=& 
2i \langle V_{0,\rm LO}^c(y)
{\cal O}_{\rm out} \rangle\,, 
\eeqa 
at leading order in continuum ChPT. 
The operator ${\cal O}_{\rm out}$ is
composed of fields with support outside the time interval 
$[y_0-t,  y_0+t]$. 
We will not need to specify ${\cal O}_{\rm out}$ in detail until
we consider the $O(a)$ corrections; for now, we only assume that
it creates an even number of pion fields, that the two-pion
component occurs at LO in a chiral expansion,
and that the two-pion component contains no zero-momentum pions.

Expanding the vector current to leading order in the pion fields, 
\bea
V_{\mu,{\rm LO}}^a &=& \frac{f^2}{2}\langle T^a( \Sigma^\dagger \partial_\mu
\Sigma +\Sigma\partial_\mu\Sigma^\dagger \rangle =
i\epsilon^{abc}\pi^b\pi_\mu^c + O(\pi^4)\,, 
\eea 
(using abbreviation $\pi_\mu^a =\partial_\mu\pi^a$), 
the right-hand-side of
the identity we want to show is simply given by 
\bea
{\rm rhs}_{\rm LO} &=&
2i \langle
V_{0,\rm LO}^c(y) {\cal O}_{\rm out} \rangle_{\rm LO}
\\ 
&=&
-2 \epsilon^{cab} \langle \pi^a(y)\pi_0^b(y) {\cal O}_{\rm out}
\rangle_{\rm LO} 
\label{rhs}
\eea 
where $\langle {\cal O}\rangle_{\rm LO}$ denotes functional
integrals with ${\cal L}_{\rm chiral}^{\rm LO}=\pi_\mu^2/2 $ in the
Boltzmann weight, and interactions, which stem from the expansion 
\bea
\label{expchirallag}
{\cal L}_{\rm chiral} &=& \frac{f^2}{4}\langle \partial_\mu \Sigma
\partial_\mu\Sigma^\dagger \rangle =\frac{1}{2}\left[ \pi_\mu^2
  +\frac{1}{3f^2}\left\{ (\pi\cdot\pi_\mu)^2 -\pi^2\,
  \pi_\mu^2\right\} \right] + O(\pi^6) \,,
\eea 
used at the lowest order giving a non-vanishing result.
In fact, to evaluate ${\rm rhs}_{\rm LO}$ we need no interactions,
given that ${\cal O}_{\rm  out}$ has a LO two-pion component.

For the left-hand-side we need the expansion of the axial-vector current, 
\bea
A_{\mu,\rm LO}^a &=& \frac{f^2}{2}\langle T^a
\left( \Sigma^\dagger \partial_\mu
\Sigma -\Sigma\partial_\mu\Sigma^\dagger\right)\rangle = i
f\left[\pi_\mu^a+\frac{2}{3f^2}\left\{\pi^a
  (\pi\cdot\pi_\mu)-\pi_\mu^a \pi^2\right\}\right]+ O(\pi^5)\,.  
\eea
The contribution with the leading order term in both axial-vector
currents on the left-hand-side vanishes, 
\bea 
-f^2\int d\vec x\langle
\left\{ \pi_0^a(y_0+t, \vec x) -\pi_0^a(y_0-t,\vec x)\right\}
\pi_0^b(y){\cal O}_{\rm out}\rangle_{\rm LO} &=& 0\,, 
\eea 
since
${\cal O}_{\rm out}$ does not contain zero momentum pion fields.

The non-vanishing contribution to the left-hand-side stems from the
three-pion term in one of the axial-vector currents or the four-pion
term in ${\cal L_{\rm chiral}}$.

We obtain the first contribution using the three-pion term in the
current $A_0^b(y)$, finding
\beqa 
{\rm lhs}_{1,\rm LO}
&=&
-\frac{2}{3}\epsilon^{abc}\int d\vec x\ \langle\left\{
\pi_0^a(y_0+t,\vec x)-\pi_0^a(y_0-t,\vec x)\right\} \left\{\pi^b
(\pi\cdot\pi_0)(y)-\pi_0^b \pi^2(y)\right\}{\cal O}_{\rm
  out}\rangle_{\rm LO}\,.\nn 
\\ 
\eea 
Performing the Wick
contractions this can be written as 
\bea 
{\rm lhs}_{1,\rm LO} &=&
2\epsilon^{cab}\langle \pi^a(y)\pi_0^b(y){\cal O}_{\rm
  out}\rangle_{\rm LO}\ \int d\vec x\ \partial_0\left\{G(t,\vec x-\vec
y)-G(-t,\vec x-\vec y)\right\}\,,\label{lhs1b} 
\eeqa 
where $G$ denotes the massless pion propagator, 
\bea
\label{pionpropcont} 
\langle \pi^a(x)\pi^b(y)\rangle_{\rm LO} \,=\, \delta^{ab} G(x-y) &=&
\delta^{ab} \int\frac{d^4 p}{(2\pi)^4} \frac{1}{p^2}e^{ip(x-y)} .
\eea 
The integral over $\vec{x}$ is easily evaluated. Using 
\beqa 
\int d\vec x\ \partial_0 G(x)&=& \left\{
\begin{array}{cc}
-\frac{1}{2} & \mbox{for}\,x_0 > 0 
\\ 
+\frac{1}{2} & \mbox{for}\,x_0 < 0 
\\
 \end{array}\right. 
\label{intd0G}
\eeqa 
the integral reduces to $-1$ and we find that
\begin{equation} 
{\rm lhs}_{1,\rm LO}={\rm rhs}_{\rm LO}\,. 
\label{eq:lhs1=rhs0}
\end{equation}

We obtain the second contribution to the left hand side using the
three-pion term for $A_0^a(y_0\pm t,\vec x)$, yielding
\bea 
{\rm lhs}_{2,\rm LO} &=& 
2\epsilon^{abc}\int d\vec x \langle \left\{ \pi^a(y_+)
\pi_0^b ( y_+) \partial_0 G(t,\vec x-\vec y) -\pi^a(y_-) \pi_0^b (
y_-) \partial_0 G(-t,\vec x-\vec y)\right\} {\cal O}_{\rm
  out}\rangle_{\rm LO}\,,\nn 
\\ & &
\label{lhs2} 
\eea 
where we
introduced the shorthand notation $y_\pm = (y_0\pm t, \vec x)$.

The third contribution stems from the four-pion terms in the chiral
Lagrangian \pref{expchirallag}, and we find
\bea 
{\rm lhs}_{3,\rm LO} &=&
\frac{1}{6}\epsilon^{abc} \int d\vec
x\ \langle\left\{\pi_0^a(y_+)-\pi_0^a(y_-)\right\} \int d^4z\left\{
(\pi\cdot\pi_\mu)^2 -\pi^2\, \pi_\mu^2\right\}\!(z)\,\, \pi_0(y) 
{\cal   O}_{\rm out}\rangle_{\rm LO}\,.  
\nn\\
& &
\eea 
This reduces to
\bea
\label{lhs3b} 
{\rm lhs}_{3,\rm LO} &=& \epsilon^{abc}\int d^4 z d\vec
x\ \langle \pi^a(z) \pi_\mu^b (z) \left\{\partial_\mu\partial_0
G(y_+-z) -\partial_\mu\partial_0 G(y_--z)\right.\nn
\\ 
&& \hspace{2.5cm}\left.+\partial_0 G(y_+-z)\partial_\mu -\partial_0
G(y_--z)\partial_\mu \right\}\partial_0 G(z-y){\cal O}_{\rm
  out}\rangle_{\rm LO}\, .  
\eea 
Since 
\beqa 
\int d\vec x \,\partial_\mu\partial_0 G(y_\pm - z) 
&=& -\delta_{\mu 0}\delta(y_0\pm t-z_0), 
\eeqa 
the first two terms in \pref{lhs3b} become 
\beqa 
&&
-\epsilon^{abc}\int d\vec x \langle \left\{ \pi^a(y_+)\pi_0^b(y_+)
\partial_0 G(t,\vec x -\vec y) - \pi^a(y_-)\pi_0^b(y_-) \partial_0
G(-t,\vec x -\vec y)\right\} {\cal O}_{\rm out}\rangle_{\rm LO}\nn \\
\label{lhs3c1}
\eeqa 
where we renamed $\vec z$ as $\vec x$.  Using \pref{intd0G} for
the integral over $\partial_0 G(y_\pm - z)$ the remaining two terms
become 
\beqa 
&& - \epsilon^{abc}\int_{y_0-t}^{y_0+t} d z_0 \int d\vec
z \langle \pi^a(z)\pi_\mu^b(z) \partial_\mu\partial_0 G(z-y){\cal
  O}_{\rm out}\rangle_{\rm LO}\,, \nn 
\eea
and after partial
integration this can be written as 
\bea 
&& -\epsilon^{abc} \int d\vec
x\langle\left\{ \pi^a(y_+)\pi_0^b(y_+)\partial_0 G(t,\vec x -\vec y)
-\pi^a(y_-)\pi_0^b(y_-)\partial_0 G(-t,\vec x -\vec y)\right\} 
{\cal O}_{\rm out}\rangle_{\rm LO}\nn 
\\ 
&& 
+\epsilon^{abc}
\int_{y_0-t}^{y_0+t} d z_0 \int d\vec z\,\langle
\pi^a(z)\partial_\mu\partial_{\mu}\pi^b(z) \partial_0 G(z-y){\cal
  O}_{\rm out}\rangle_{\rm LO}\,.\nn 
\eeqa 
Since
$\partial_\mu^2\pi^b(z)$ is contracted with the on-shell fields in
${\cal O}_{\rm out}$ (recall our assumption!), the second line
vanishes.  The remaining first line is the same as in
\pref{lhs3c1}. Hence, in total the third contribution reduces to 
\beqa
     {\rm lhs}_{3,\rm LO} &=& - {\rm lhs}_{2,\rm LO}\,,
\label{lhs3d}
\eeqa 
so the second and third contributions cancel.
This, together with eq.~\pref{eq:lhs1=rhs0}, 
proves the Ward-Takahashi identity \pref{WTIgen} to first
non-trivial order in the chiral expansion.

Repeating this calculation with the lattice spacing corrections
included is now straightforward. What changes are the expansions of
the currents and the effective Lagrangian in terms of the pion fields:
\bea 
\Lagno_{\rm chiral} & = & \frac{1}{2} (1 + X_1) \pimu^2 +
\frac{1}{6f^2} \left[(1+X_1)(\pi\cdot\pimu)^2 -
  (1+X_2)\pi^2\pimu^2\right] + O(\pi^6)\,,
\\ 
V_{\mu,\rm eff}^a & = &
i\epsilon^{abc}\pi^b\pimu^c (1+X_1) + O(\pi^4)\,,
\\
A_{\mu,\rm eff}^a &=& if
(1+Y_1) \pimu^a + i \frac{2}{3f}\left[(1+Y_2)\pi^a(\pi\cdot\pimu)
  -(1+Y_3)\pimu^a \pi^2\right]\,, 
\eea 
where we introduced the
coefficients 
\bea 
X_1 &=& \frac{16}{f^2} W_{45}\hat{a}\csw\,,\nn
\\
X_2
&=& \frac{40}{f^2} W_{45}\hat{a}\csw\,,\nn
\\ 
Y_1 &=& X_1+
\frac{8}{f^2} W_{A}\hat{a}\ca\,,\nn
\\ 
Y_2 &=& X_1- \frac{4}{f^2} W_{A}\hat{a}\ca\,,\nn
\\ 
Y_3 &=& \frac{28}{ f^2} W_{45}\hat{a}\csw +
\frac{2}{f^2} W_{A}\hat{a}\ca\,.\label{XY} 
\eea 
The modification of
the kinetic term by the factor $1+X_1$ yields a slightly different
pion propagator, 
\bea
\label{pionpropa} 
\langle
\pi^a(x)\pi^b(y)\rangle_{{\rm LO}(a)} \,=\, \frac{\delta^{ab}}{1+X_1}
\langle \pi^a(x)\pi^b(y)\rangle_{\rm LO}\,=\,
\frac{\delta^{ab}}{1+X_1} G(x-y) \,, 
\eea 
with $G$ defined in \pref{pionpropcont} and 
$\langle {\cal O}\rangle_{{\rm LO}(a)}$
denotes functional integrals with 
${\cal L}_{\rm chiral}^{\rm LO}=(1+X_1)\pi_\mu^2/2 $ 
in the Boltzmann weight. Notice that
$1+X_1=Z_{\pi}^{-1}$ [cf.\ \pref{eq:Zpi}]. Hence, in terms of the
renormalized pion fields 
\bea 
\tilde{\pi}^a(x) &=& Z_{\pi}^{-1/2} \pi^a(x)\,, 
\eea 
equation \pref{pionpropa} assumes its standard form
\bea
\label{pionproprenfield} 
\langle
\tilde{\pi}^a(x)\tilde{\pi}^b(y)\rangle_{{\rm LO}(a)} \,=\,
\delta^{ab}G(x-y) \,, 
\eea 

The right hand side of the WTI now reads
\bea
\label{rhsincla} 
{\rm rhs}_{\rm LO(a)}&=&
-2 \epsilon^{cab} (1+X_1) \langle
\pi^a(y)\pimu^b(y) {\cal O}_{\rm out} \rangle_{{\rm LO}(a)} \,.
\eea 
while the three contributions on the left hand side are 
modified as follows:
\bea 
{\rm lhs}_{1,\rm LO(a)} &= & 
\frac{\left[1 + (3 Y_1 + Y_2 + 2Y_3)/3\right]}{(1+X_1)} 
\ (-2 \epsilon^{cab}) \langle
\pi^a(y)\pimu^b(y) {\cal O}_{\rm out} \rangle_{{\rm LO}(a)}
\,,\label{lhs1a}
\\ 
{\rm lhs}_{2,\rm LO(a)} &=& 
\frac{\left[1 + (3 Y_1 + Y_2 + 2Y_3)/3\right]}{(1+X_1)} 
\left({\rm lhs}_{2,+} - {\rm lhs}_{2,-}\right)
\,,\label{lhs2a}
\\
{\rm lhs}_{2,\pm} &=& 
2\epsilon^{abc}
\int d\vec x \, \partial_0 G(\pm t,\vec x-\vec y) 
\langle \pi^a(y_\pm) \pi_0^b (y_\pm) 
{\cal O}_{\rm out}\rangle_{{\rm LO}(a)} 
\,,\label{lhs2pma}
\\ 
{\rm lhs}_{3,\rm LO(a)} &= &
- \frac{\left[1 + 2Y_1 + ( X_1 + 2X_2)/3\right]}{(1+X_1)^2}
\left({\rm lhs}_{2,+} - {\rm lhs}_{2,-}\right)\,,
\label{lhs3a} 
\eea 
Here the factors in the numerator come from
\pref{XY} while those in the denominator arise from the
presence of (one or two) pion propagators in the manipulations
given earlier in the appendix.

To determine the renormalization constant we need the ratio
of the two sides of the original WTI. It follows from
\pref{rhsincla} and \pref{lhs1a} that
\begin{equation}
\frac{{\rm lhs}_{1,\rm LO(a)}}{{\rm rhs}_{\rm LO(a)}} =
\frac{\left[1 + (3 Y_1 + Y_2 + 2Y_3)/3\right]}{(1+X_1)^2} 
\,,
\end{equation}
independent of the detailed form of ${\cal O}_{\rm out}$.

For the remaining ratios we
need to specify the external fields.  We take 
\bea 
{\cal O}^c_{\rm out} & =& 
\epsilon^{cde} \tilde{\pi}^d(T,\vec{p} ) \tilde{\pi}^e(-T,-\vec q)
\eea 
with $\vec p,\ \vec q\not= \vec 0$. Here $\tilde{\pi}^d(T,\vec{p})$ is the
Fourier transform of $\tilde{\pi}^d(T,\vec{x})$ with respect to the
three spatial coordinates.  
For this choice, the correlators we need are
\beqa 
{\rm rhs}_{\rm LO(a)} &=& -12\ 
 \partial_0^T \{G(T-y_0,\vec p) G(T+y_0,\vec q)\}
\,,\\
{\rm lhs}_{2,\pm} &=& 12\ 
\frac{\partial_0 G(\pm t, \vec p-\vec q)}{(1+X_1)} 
\partial_0^T \{G(T-y_0\mp t,\vec p) G(T+y_0 \pm t,\vec q)\}\,,
\eeqa 
where $\partial_0^T= \partial/\partial T$.
For simplicity, and without loss of generality,
we have set $\vec y=0$, which avoids an overall phase.

To evaluate these expressions we need the hybrid position-momentum
pion propagator,
\beqa
\label{Gt} G(t,\vec p)\, =\,\int 
\frac{d p_0}{2\pi}\frac{e^{i p_0 t}}{p_0^2 +\vec{p}^2} \, =\, 
\frac{e^{-E_p t}}{2E_p}, \quad t\ge 0,\, E_p =\vert \vec p\vert, 
\eeqa 
using which we find
\bea 
{\rm rhs}_{\rm LO(a)}
&=& 
\frac{3(E_p+E_q)}{E_pE_q}
\ e^{\left[-(E_p+E_q) T+ y_0 (E_p-E_q)\right]}\,,  
\\
{\rm lhs}_{2,\pm} &=&
\frac{\pm1}{(1+X_1)} \frac{3(E_p+E_q)}{2E_pE_q}
\ e^{\left[-|t| E_{p-q} -(E_p+E_q) T+ (y_0 \pm t) (E_p-E_q) \right]}\,,
\eea 
where $E_{p-q}=|\vec p-\vec q|$.
Inserting these results into
eqs.~\pref{rhsincla}, \pref{lhs2a} and \pref{lhs3a}
we obtain
\bea
\frac{{\rm lhs}_{2,\rm LO(a)}}{{\rm rhs}_{\rm LO(a)}} &=&
\frac{\left[1 + (3 Y_1 + Y_2 + 2Y_3)/3\right]}{(1+X_1)^2} 
\ 
\cosh[t(|\vec p|-|\vec q|)] 
\exp[-|t| |\vec p - \vec q|]
\,,\\
\frac{{\rm lhs}_{3,\rm LO(a)}}{{\rm rhs}_{\rm LO(a)}} &=& -
\frac{\left[1 + 2Y_1 + ( X_1 + 2X_2)/3\right]}{(1+X_1)^3}
\
\cosh[t(|\vec p|-|\vec q|)] 
\exp[-|t| |\vec p - \vec q|]
\,,
\eea
and thus, finally,
\bea
\label{eq:lhs_over_rhs_final}
\frac{{\rm lhs}_{\rm LO(a)}}{{\rm rhs}_{\rm LO(a)}} 
&=& 
\frac{{\rm lhs}_{1,\rm LO(a)}+{\rm lhs}_{2,\rm LO(a)}
+{\rm lhs}_{3,\rm LO(a)}}{{\rm rhs}_{\rm LO(a)}} 
\\
&=& 
1 + 
\frac{8\hat a}{f^2}(W_{45}\csw+W_A\ca)
\left\{1 - 
\cosh[t(|\vec p|-|\vec q|)] 
\exp[-|t| |\vec p - \vec q|]
\right\} 
\,.
\eea
This ratio should be unity when
multiplied by the renormalization factor
$(Z_{\rm A,Loc}/Z_A^0)^2$,
leading to the result \pref{ZAWChPT} in the main text.

\end{appendix}

\end{document}